\documentclass[journal]{IEEEtran}
\ifCLASSINFOpdf
\else
\fi
\usepackage{times}
\usepackage{fancyhdr,graphicx,amsmath,amssymb}
\usepackage[ruled,vlined]{algorithm2e}
\usepackage{empheq} 
\usepackage{cases} 
\usepackage{verbatim}
\usepackage{fdsymbol}
\usepackage{url}
\usepackage[hyphenbreaks]{breakurl}

\usepackage{xcolor}
\usepackage{multirow}

\usepackage{booktabs,caption}
\usepackage[flushleft]{threeparttable}
\usepackage{rotating}
\usepackage{cite}

\usepackage{tabulary}

\usepackage{dblfloatfix}
\usepackage{amssymb}
\usepackage{pifont}
\newcommand{\cmark}{\ding{51}}%

\begin{document}
%



\title{A Survey on Offloading in Federated Cloud-Edge-Fog Systems with Traditional Optimization and Machine Learning}
%
%
%

\author{Binayak~Kar,~\IEEEmembership{Member,~IEEE,}
        Widhi Yahya,
        Ying-Dar~Lin,~\IEEEmembership{Fellow,~IEEE,}
        and Asad Ali,~\IEEEmembership{Student Member,~IEEE}

\IEEEcompsocitemizethanks{
\IEEEcompsocthanksitem B. Kar is with the Department of Computer Science and Information Engineering, National Taiwan University of Science and Technology, Taipei, 106, Taiwan. E-mail: bkar@mail.ntust.edu.tw.
\IEEEcompsocthanksitem W. Yahya, and A. Ali are with the Department of Electrical Engineering and Computer Science, National Yang Ming Chiao Tung University, Hsinchu, 300, Taiwan. E-mail: widhi.yahya@ub.ac.id; ali.eed06g@nctu.edu.tw
\IEEEcompsocthanksitem Y.-D Lin is with the Department of Computer Science, National Yang Ming Chiao Tung University, Hsnchu, 300, Taiwan. E-mail: ydlin@cs.nctu.edu.tw.
}
}

\maketitle

\begin{abstract}
The huge amount of data generated by the Internet of things (IoT) devices needs the computational power and storage capacity provided by  cloud, edge, and fog computing paradigms. Each of these computing paradigms has its own pros and cons. Cloud computing provides enhanced data storage and computing power but causes high communication latency. Edge and fog computing provide similar services with lower latency but with limited capacity, capability, and coverage. A single computing paradigm cannot fulfil all the requirements of IoT devices and a federation between them is needed to extend their capacity, capability, and services. This federation is beneficial to both subscribers and providers and also reveals research issues in traffic offloading between clouds, edges, and fogs. Optimization has traditionally been used to solve the problem of traffic offloading. However, in such a complex federated system, traditional optimization cannot keep up with the strict latency requirements of decision making, ranging from milliseconds to sub-seconds. Machine learning approaches, especially reinforcement learning, are consequently becoming popular because they can quickly solve offloading problems in dynamic environments with large amounts of unknown information. This study provides a novel federal classification between cloud, edge, and fog and presents a comprehensive research roadmap on offloading for different federated scenarios. We survey the relevant literature on the various optimization approaches used to solve this offloading problem, and compare their salient features.  We then provide a comprehensive survey on offloading in federated systems with machine learning approaches and the lessons learned as a result of these surveys. Finally, we outline several directions for future research and challenges that have to be faced in order to achieve such a federation.

\end{abstract}


%
\IEEEpeerreviewmaketitle

\section{Introduction}
%
%
%
%
\IEEEPARstart{T}{here} 
are many computing paradigms which provide computational power and storage services for the huge amounts of data generated by an ever-increasing number of heterogeneous devices. Three of the most well-known and widely adopted computing paradigms are cloud, edge, and fog computing. 
The terms \textit{cloud}, \textit{edge}, and \textit{fog} represent three computing tiers of  cloud, edge, and fog computing systems. 

\begin{figure}[!t]
\centerline{\includegraphics[width=18.5pc]{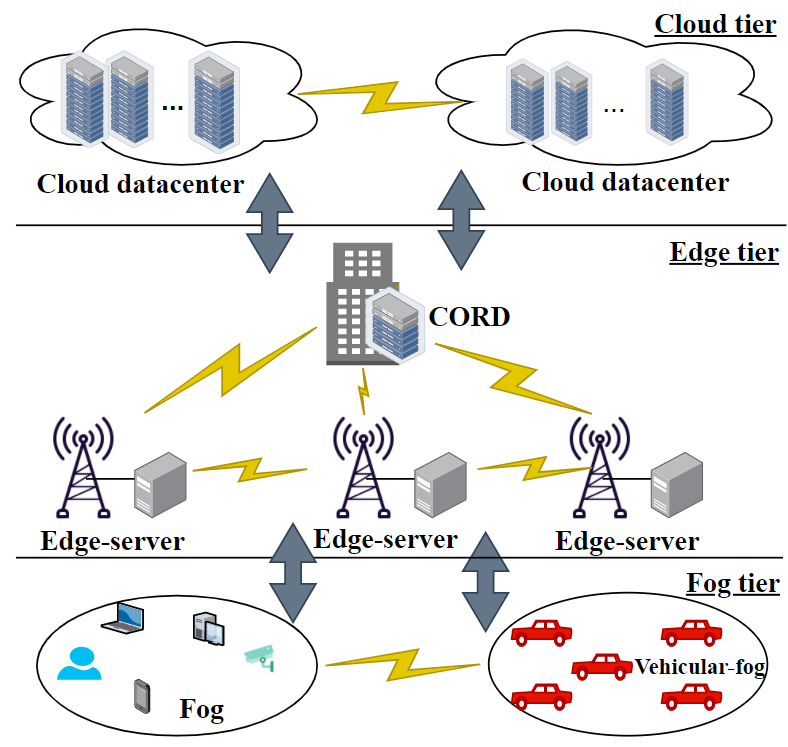}}
\caption{Integrated cloud, edge, and fog systems}
\label{fig:model}
\end{figure}

Figure~\ref{fig:model} shows the cloud-edge-fog system consists of three tiers. 
(a) Cloud tier: the top tier is a cloud system \cite{B01}, 
that encompasses the cloud computing paradigm which is the most well-known and widely adopted computing paradigm for more than a decade because of its attractive features such as scalability, rapid elasticity, resource pooling, cost saving, and easy maintenance. This tier consists of different clouds, such as Google and Amazon. These basically deal with industrial big data, business logic, and analytical databases, data “warehousing,” and so on.  
(b) Edge tier: the middle tier is an edge system \cite{B02}, 
that comprises the edge computing paradigm. Edge computing has its origins with European Telecommunication Standards Institute (ETSI) that proposed virtualizing the capabilities of cloud computing into mobile network operators. An edge server can be deployed behind a cellular system’s base station and central office. This tier includes different service providers such as Verizon, T-Mobile, AT\&T, Chunghwa telecom, and so on, and consists of local network assets, micro-data centers, central offices, base stations, etc. 
(c) Fog tier: the bottom tier is a fog system \cite{B03} or IoT system consisting of both mobile users (e.g., smartphones, tablets, and laptops), smart vehicles forming vehicular-fogs \cite{B04}, and IoT devices, such as industrial actuators, wearable devices, and smart sensors,  and the like. It covers real-time data processing on industrial PCs, process-specific applications and autonomous equipment, group of local computing devices, electronic vehicles, etc.

\subsection{The Federation: Cloud, Edge, and Fog}

The Internet of Things (IoT) devices that have taken the world by storm need computational power and storage capacity for the huge amount of data generated by them so as to provide services to their subscribers \cite{B1}. 
Cloud, edge, and fog computing are the potential paradigms that could fulfil the demand of subscribers \cite{B2}. 
Cloud computing is the on-demand availability of computer system resources, especially data storage and computing power, without the need for direct active management by a user \cite{B01}. However, cloud computing introduces high communication latency as  its servers are far from  end-users or subscribers. A cloud computing paradigm is not suitable for some applications with very strict communication latency limitations. This is where edge and fog computing models play a crucial role in providing similar services with lower latency \cite{B4}\cite{B5}.\par

Again, all these computing paradigms, \textit{i.e.}, cloud, edge, and fog, have their own limitations in the terms of capacity, capability, coverage, storage, and latency. A single computing paradigm cannot by itself fulfil the diverse requirements of a vast number of traditional and heterogeneous IoT devices. For example, a user might need to use two different applications at the same time and one of them is latency sensitive while the other is computation sensitive. In this case, the user would require the services provided by both cloud and edge or fog \cite{SS13}.
Also, if a cloud customer needs some extra service that is not available in that cloud, then the cloud must try to arrange that service for the customer without a delay as to provide satisfaction. The cloud may otherwise lose the trust of the customer and, in some cases, may lose the customer, which may affect its business financially and in reputation. This is where a federation between multiple computing paradigms  can play a key role in resolving these issues. Such a federation is not only suitable for subscribers but also for providers. A subscriber will be able to access the services provided by different computing paradigms without having to buy multiple subscriptions. On the other hand, providers would be able to extend their capacity, capability, and coverage without having to lose  subscribers to other providers. \par
\subsection{Offloading in Federated Systems}
A federation between multiple computing paradigms gives rise to many different opportunities and challenges such as authentication, access control, resource sharing, and traffic offloading. In this work, we focus on  offloading in a federated environment where cloud, edge, and fog offload  traffic to each other. Such offloading is basically a transfer of tasks that are resource intensive to a separate platform in order to perform a task in a better way. Such  offloading becomes necessary when a task assigned to a service provider exceeds its computing resources and has to be offloaded to another service provider that can provide the required computing power. Offloading is thus required in order to fulfil different constraints under different situations. Some of the important constraints to be overcome are latency, load balancing, privacy, and storage constraints.

There can be two types of offloading In a federated system, \textit{intra-domain} and \textit{inter-domain} offloading, as there are multiple domains in such a federation. Intra-domain offloading involves the traffic offloaded between the entities belonging to the same tier \textit{i.e.,} cloud-to-cloud, edge-to-edge, or fog-to-fog, while inter-domain offloading involves the entities belonging to different tiers, such as cloud-to-edge, cloud-to-fog, or edge-to-fog etc. \par

Optimization has traditionally been used effectively to offload traffic in single networks \cite{B5a} or in a federation as a single network provides the optimal offloading ratio that will reduce the overall cost of the network. Although traditional optimization has been used for years, it takes much time to generate decisions because of a network’s complexity and the very large number of variables involved. The non-convex algorithms in traditional optimization perform an exhaustive search to find an optimal solution, and which takes much time to converge \cite{ex1}. Modern applications are latency sensitive and cannot afford such delays in offloading decisions, as the control and data planes need a decision in milliseconds to subseconds. In the current era, optimization solutions for quick offloading decisions are becoming obsolete and machine learning approaches are taking the place of traditional optimization in complex network systems because of their faster response times. 

The machine learning approach has an advantage over the traditional optimization approach in such complicated federated systems because machine learning does not require a complete knowledge of the system compared to traditional optimization, and it can quickly solve  offloading problems with various bits of unknown information. In the various machine learning approaches, reinforcement learning (RL) is the most suitable for offloading decisions because RL does not need a  data set and  can learn directly from the environment  \cite{SS16}. This makes RL suitable for offloading decisions in a dynamic environment with much unknown information. This also shows that RL is better than the traditional optimization approach because in such complex systems traditional optimization may not be able to converge to an optimal solution and may preferably rely on heuristics. Traditional optimization would take much more time for decision making  compared to RL because of exhaustive searching. When we consider offloading in a complex federated environment together with traditional optimization, machine learning, and reinforcement learning, many research opportunities and challenges arise. 

We summarize these various research opportunities such as V2X, fog-fog federation, mobility in vehicular fog, scaling, resource allocation, centralized vs. distributed federation, etc. below. We also provide some insight into the important challenges that will need to be faced by the operators to deploy this kind of federation such as redundancy, fault tolerance, SLA, reliability, geo-diversity, performance, security, and interoperability between entities of the different domains in a federated environment. \par

\begin{table*}[!t]
\renewcommand{\arraystretch}{1.25}
\caption{Survey on Surveys on Offloading in the Federated Systems}
\label{table-0}
\centering
\begin{tabular}{|c|c|c|c|c|c|c|c|l|}
\hline
 & \multicolumn{5}{l|}{\textbf{Coverage}} &  &  &\\ \cline{2-6} 
\textbf{\multirow{2}{*}{References}}  & \begin{turn}{-90} \textbf{Cloud} \end{turn}& \begin{turn}{-90} \textbf{Edge} \end{turn} & \begin{turn}{-90} \textbf{Fog} \end{turn} & \begin{turn}{-90} \textbf{Vehicular-fog} \end{turn} & \begin{turn}{-90}\textbf{Device}\end{turn} &\begin{turn}{-90} \textbf{ Federation model\;}\end{turn} &  \begin{turn}{-90} \textbf{ Optimization}\end{turn} &  \textbf{\multirow{2}{*}{Focus of the survey}}\\ \hline

\cite{SS2}                   &                            &                           &                          &                                    & \cmark                           &                                     &                                &  Data offloading techniques in cellular networks                          \\ \cline{1-1} \cline{6-6} \cline{9-9} 
\cite{SS3}                                          &                            &                           &                          &                                    & \cmark                           &                                     &                                &  Offload techniques and   management in wireless access and core networks \\ \cline{1-1} \cline{6-6} \cline{9-9} 
\cite{SS4}                                          &                            &                           &                          &                                    & \cmark                           &                                     &                                &  Traffic   offloading in heterogeneous cellular network                    \\ \cline{1-1} \cline{6-6} \cline{9-9} 
\cite{SS9}                                             &                            &                           &                          &                                    & \cmark                           &                                     &                                &  Opportunistic Offloading                                                 \\ \cline{1-1} \cline{6-6} \cline{9-9} 
\cite{SS11}                                           & \multirow{-5}{*}{}        &                           &                          &                                    & \cmark                           &                                     &                                &  Mobile   data offloading technologies                                    \\ \cline{1-2} \cline{6-6} \cline{9-9} 
\cite{SS1}                                        &                            &                           &                          &                                    & \cmark                           &                                     &                                &  Computation offloading for mobile systems                                \\ \cline{1-1} \cline{6-6} \cline{9-9} 
\cite{SS5}                                        &                            &                           &                          &                                    & \cmark                           &                                     &                                &  Security   and privacy challenges in mobile cloud computing              \\ \cline{1-1} \cline{6-6} \cline{9-9} 
\cite{SS7}                                   &                            & \multirow{-8}{*}{}       &                          &                                    & \cmark                           &                                     &                                &  Adaptation   techniques in computation offloading                        \\ \cline{1-1} \cline{3-3} \cline{6-6} \cline{9-9} 
\cite{SS6}                                           &                            &                           &                          &                                    &                            &       Single                              &                                &  Architecture   and Computation Offloading                                \\ \cline{1-1} \cline{6-6} \cline{9-9} 
\cite{SS10}                                            &                            &                           &                          &                                    &                            &                                     &                                &  Multi-objective   decision-making for mobile cloud offloading            \\ \cline{1-1} \cline{6-6} \cline{9-9} 
\cite{SS14}                                           &                            &                           &                          & \multirow{-11}{*}{}               & \cmark                           &                                     &                                &  A Survey and taxonomy on task offloading for   edge-cloud computing      \\ \cline{1-1} \cline{5-6} \cline{9-9} 
\cite{SS15}                                             &                            &                           & \multirow{-12}{*}{}     & \cmark                                  &                            &                                     &                                &  Computation   offloading for vehicular environments                      \\ \cline{1-1} \cline{4-6} \cline{9-9} 
\cite{SS13}                                         &                            &                           &                          &                                    &                            &                                     &                                &  Computation offloading in edge-cloud systems                             \\ \cline{1-1} \cline{6-6} \cline{9-9} 
\cite{SS8}                                         &                            &                           &                          &                                    & \cmark                           &                                     &                                &  Offloading in fog computing: Enabling technologies                       \\ \cline{1-1} \cline{6-6} \cline{9-9} 
\cite{SS12}                                            &                            &                           &                          &                                    & \cmark                           &                                     &                                &  Data offloading techniques in V2X networks                               \\ \cline{1-1} \cline{6-6} \cline{9-9} 
\cite{SS17}                                             &                            &                           &                          &                                    & \cmark                           &                                     & \multirow{-16}{*}{T}           &  Computation   offloading modeling for edge computing                     \\ \cline{1-1} \cline{6-6} \cline{8-9} 
\cite{SS16}                                      &                            &                           &                          & \multirow{-5}{*}{}                & \cmark                           & \multirow{-17}{*}{}                & ML                             &  Machine learning-based approaches in mobile edge computing               \\ \cline{1-1} \cline{5-9} 
Our                                               & \multirow{-13}{*}{\cmark}       & \multirow{-10}{*}{\cmark}      & \multirow{-6}{*}{\cmark}      &                                 \cmark  & \cmark                           &                                    Multiple & T/ML                           &  Offloading in federated cloud-edge-fog systems                                                                                               \\ \hline

\multicolumn{9}{l}{
    \begin{minipage}{6.5cm}
          \small T: Traditional; ML: Machine Learning
    \end{minipage}
}
\end{tabular}

\end{table*}

\subsection{Overview of Surveys}
In this section, we discuss recent works that survey offloading in federated systems, as well as the importance of our survey. Table \ref{table-0} compares offloading surveys which are divided into coverage, federation models, optimization approaches, and what the focus of that survey is.  

The authors of \cite{SS2,SS3,SS4, SS9, SS11} discussed traffic and data offloading between cellular, WiFi, and opportunistic networks, but did not consider computation offloading in a federated system such as an edge-cloud system. Rebecchi et al. \cite{SS2} reviewed data offloading approaches in  cellular system with WiFi environment and categorized them based on their latency requirements. Maallawi et al. \cite{SS3} surveyed offloading and management approaches in wireless access and in core networks. Their objective was to address providers' problems such as radio access scheduling, revenue per user decrease, and coverage. Chen et al. \cite{SS4} surveyed traffic offloading in heterogeneous cellular networks, including small cells, WiFi networks, and opportunistic networks, and \cite{SS9} focused on the algorithm for selecting the optimal subset of offloading nodes in an opportunistic network, which would allow a node to offload traffic and computation tasks to another node. This kind of D2D offloading is beneficial to a cellular operator and users in terms of monetary cost. Huan et al. \cite{SS11} surveyed  mobile data offloading, which involves small cells, WiFi networks, opportunistic networks, and heterogeneous networks. The pros and cons of each of these networks are also detailed.

Computation offloading between mobile device and cloud is discussed in \cite{SS1, SS5, SS7}. Kumar et al. \cite{SS1} categorized offloading techniques based on the decision characteristics and applications. The security and privacy challenges in mobile cloud computing are discussed in \cite{SS5}. The offloading techniques with environmental variation which included applications, networks, execution platforms, and cloud managements are summarized in \cite{SS7}. 

Edge-cloud system offloading was surveyed in \cite{SS6, SS10, SS14, SS15, SS13, SS8, SS12, SS17, SS16}. Mach et al. \cite{SS6} discussed mobile edge-cloud system architectures and considered computation offloading, resources allocation and mobility management. Wu et al. \cite{SS10} discussed multi-objective offloading, which was initiated by a large heterogeneous system such as mobile edge computing. Response time and energy consumption were their two main objectives. Offloading criteria were categorized into what, when, where, and how to offload. The taxonomy of edge-cloud offloading was categorized in \cite{SS14}, based on task type, offloading scheme, objectives, device mobility, and multi-hop cooperation. De et al. \cite{SS15} presented a classified taxonomy of V2X system offloading, based on a communication standard, problem, and experiment.

Fog-edge-cloud offloading was discussed in \cite{SS13, SS8, SS12, SS17, SS16}. Jiang et al. \cite{SS13} surveyed and discussed state-of-the-art computational offloading in mobile edge computing. Aazam et al. \cite{SS8} discussed the offloading technologies in fog computing for IoT. The survey of Zhou et al. \cite{SS12} focused on vehicular offloading, which included vehicle-to-vehicle, vehicle-to-infrastructure, and vehicle-to-everything, with a brief discussion of the architecture design, algorithm, and problem formulation. Lin et al. \cite{SS17} focused on offloading modeling, which included communication, computation, energy harvesting, and channel modeling. Shakarami et al. \cite{SS16} classified  machine learning-based offloading into approaches such as supervised ML, unsupervised ML, and reinforcement learning~(RL). 

None of these surveys details a federation between cloud, edge, fog, and vehicular fog. Each combination of such a federation has different characteristics and offloading directions which leads to complex issues. In most of the surveys traditional optimization was used to optimize the offloading decision in a federation. Traditional optimization takes a long time to converge in such a complex federation system. By contrast, an offloading decision must be rapidly determined by the control plane. ML-based approaches have recently become popular to solve offloading optimization problems in such a complex federation system with fast response times. This survey focuses on edge-cloud federation offloading and covers state-of-the-art offloading approaches that use ML.

\subsection{Contribution} 

The major contributions in this paper are as follows. First, we discuss the classification of federation between cloud, edge, and fog systems, followed by a classification the possible offloading techniques between the entities of a federated system. We discuss the current research status of different federated architectures and offloading techniques, and survey  offloading based on traditional optimization and machine learning approaches and make a comparative study of both approaches. Finally, we discuss some key research challenges associated with the task offloading and point to possible future research directions.

The rest of this paper is organized as follows. Section II describes a federation of fog-edge-cloud and classification of such a federation. Section III presents the offloading, classification of the offloading, and the current research status of federated architecture and offloading. The survey on offloading is detailed in Section IV, which also classified the approaches into traditional optimization and machine learning. Lessons learned from the survey are discussed in Section V. The research opportunities and challenges are presented in Section VI and the conclusions of this survey discussed in Section VII.


\section{A Federation}

A federation can be defined as the collection of clouds that cooperate to provide resources requested by users  \cite{B10}. Stated another way, a cloud can provide computing resources wholesale or rent to another cloud provider \cite{B11}. A federation can render the cloud a user and resource provider at the same time \cite{B12}. A customer’s request submitted to one cloud can be fulfilled by another cloud. A cloud provides capacity and coverage, but for latency reduction and fault tolerance, a cloud needs edge and fog. Likewise, fog and edge need the service of a cloud to increase their capacity and coverage. 

Cloud, edge, and fog computing paradigms provide different services to users or subscribers, depending on their limitations and capacity  \cite{B13}. Since subscribers have different demand and service requirements, each paradigm may not have all kinds of services to fulfil all users’ needs,  because each computing paradigm has its limitations \cite{B14}. There is thus a need for a “federation” between different service providers to cope with a users’ heterogeneous requirements and increase the capacity, coverage, capability, and fault tolerance of  service providers. For example, in a smart city environment, different users or IoT devices have different requirements that may not be addressed by a single service provider. A federation then comes into the picture to fulfil the various demands. The benefits of the federation are twofold, i.e., from both a subscriber’s and provider’s perspectives. A subscriber would not have to subscribe to the services of all providers but will get the services of all by  just  subscribing to one of them. Subscribers do not have to keep multiple accounts and do not have to pay  multiple providers. On the other hand, a provider will not lose a customer just because it cannot provide a particular service.

\begin{figure*}[hbt!]
\centerline{\includegraphics[width=14cm,height=11cm,keepaspectratio]{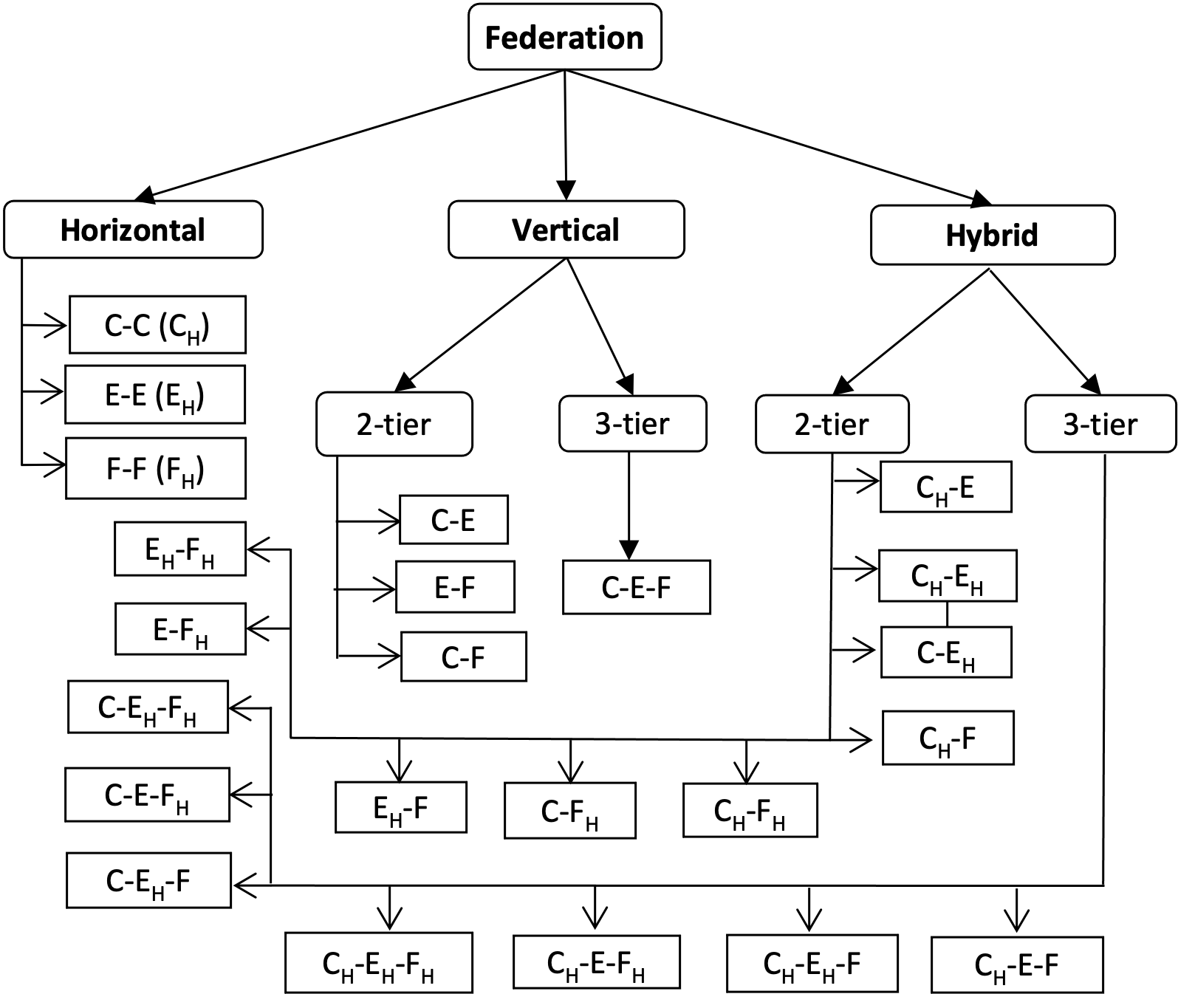}}
\caption{Classification of federation}
\label{fig:federation}
\end{figure*}

\subsection{Federation vs. Non-federation}
A non-federated scenario is one where a service provider cannot share its resources with other service providers, and it can neither lend nor rent its surplus resources to others. In such a scenario, it is difficult to handle the dynamic demands of users, and the service provider may face issues like Lock-in  \cite{O1}\cite{O2}\cite{O3} and single point failure \cite{A1}\cite{A2}. 
Lock-in is one of the most cited and controversial obstacles to widespread cloud computing adopted by enterprises \cite{O4}.
It is also risky for a customer to be tied to a single vendor because that vendor might raise prices, go out of business, become unreliable, or fail to keep up with technological progress. 

Different service providers, such as cloud, edge, or fog, provide different services to their subscribers depending on their limitations and capacity \cite{B9}. 
Again, a subscriber of a service provider may have different demands at a different time, which may not be fulfilled by the service provider always. This can also be understood with the help of an example of a perfect smart city where there would be different types of IoT devices, and each type would have its own requirements that a single IoT service provider cannot fulfil. In such a scenario, the provider will be able to provide all sorts of IoT services after federating with other providers. When acquiring IoT deployment, a federated environment is thus more beneficial than a non-federated scenario.

\subsection{Classification of a Federation}
With federation technology, different users of or subscribers to different service providers get different benefits. With this technology, different service providers can federate with each other to provide a better service to their users. A federation between these service providers can be divided into three categories, horizontal, vertical, and hybrid federations. These federations are all based on the cloud, edge and fog integrated architecture shown in  Figure~\ref{fig:model}; the classification of all possible federation scenarios is shown in Figure~\ref{fig:federation}. To the best of our knowledge, such a classification has not been dealt with in the any of the  studies we reviewed. 
%

\textit{1) Horizontal Federation.} A horizontal federation consists two federated entities in the same tier, such as a cloud-cloud federation \cite{A3}. A horizontal federation can be  cloud-cloud $(C-C)$ or $C_H$, edge-edge $(E-E)$ or $E_H$, or Fog-Fog $(F-F)$ or $F_H$. 

\textit{2) Vertical Federation.}
A vertical federation is a federation between entities in different tiers \cite{B12} as in a cloud-edge federation. Since a cloud-edge-fog system is a three-tier system, we can classify a vertical federation into two and three-tier federations, such as cloud-edge  $(C-E)$, edge-fog $(E-F)$, and cloud-fog $(C-F)$ federations, or a cloud-edge-fog $(C-E-F)$ federation. 

\textit{3) Hybrid Federation.}
A hybrid federation is a federation that combines both horizontal and vertical combinations  \cite{B20}, where entities can simultaneously federate horizontally with another entity in the same tier as well as can vertically with entity in another tier. For example, in an edge-edge-cloud $(E_H-C)$ federation, an edge is federated with another edge in tier-2 and also federated with a cloud in tier-3. Such a hybrid federation can be classified into two-tier and three-tier federations. 

\begin{enumerate}
\item A two-tier hybrid federation is the combination of
all possible combinations of horizontal and two-tier vertical federations. For example, in $(C_H-E)$ federation, one cloud $(C_1)$ will federate with another cloud $(C_2)$ horizontally and with an edge $(E_1)$ vertically. Similarly, in  $(C_H-E_H)$, two clouds $(C_1~ and ~C_2)$ become federated with each other, two edges $(E_1~ and~ E_2)$ are federated with each other horizontally, and are simultaneously also federated vertically $(E_1~ and~ C_1)$. All nine possible two-tier hybrid federation  combinations are shown in Figure~\ref{fig:federation}.

\item A three-tier hybrid federation is the combination of all possible combinations of a horizontal federation and three-tier vertical federation; all seven possible federation combinations are also shown in Figure~\ref{fig:federation}. For example, in  $(C_H-E_H-F_H)$, two clouds $(C_1~ and ~C_2)$ in tier-3, are federated with each other, two edges $(E_1~ and~ E_2)$ in tier-2 are federated with each other, and  two fogs $(F_1~ and~ F_2)$ in tier-2 get federated with each other. At the same time, $C_1$ with $ E_1$ and $ E_1$ with $ F_1$ also become federated vertically.
\end{enumerate}

\section{Offloading}
When an entity or service provider (say SP1) with a federated architecture receives requests from its subscribers or customers, and needs another entity (say SP2) to execute tasks on behalf of SP1 and return the results. This is called offloading \cite{A4}. Again, there are various criteria that are used when deciding whether or not to offload certain tasks. A few examples of this are as follows. To meet a resource constraint: when a task requires more computing resources than the local system’s available capacity, it must be offloaded to another system with the required capacity \cite{A5}. To address latency: as distance affects time-sensitive applications, the node closest to the receiving node must be involved in the task of offloading so as to provide the services faster \cite{A6}. Load balancing: when a server has reached its capacity for executing tasks, additional tasks need to be distributed between other entities in the service provider’s ecosystem \cite{A7}. Storage: small computing devices with limited storage facilities may require offloading to another that has  large storage capacity \cite{A8}. To maintain privacy, confidentiality, and security: depending on the sensitive of data, they may be offloaded to more secure cloud storage \cite{A9}.


\subsection{Renting vs. Scaling vs. Offloading}  
In an offloading scenario, resources are used based on the requests from customers. These may vary from time to time, based on demand. Here the use of resources can be scaled up or down based on the demand, and a customer will pay according to use. This is called autoscaling. However, in renting, a customer will reserve the required resources for a predetermined duration for future use. The customer may or may not utilize the entire resources that was reserved, but will pay according to the reservation. Offloading is a method where a service provider passes the request to another service provider to provide the service to its own customer. For example, a client of Amazon sends a request to Amazon, but Amazon passes the request to Google, and Google provides the service, providing there is a federation agreement between the two service providers.

\begin{figure*}[hbt!]
\centerline{\includegraphics[width=16cm,height=13cm,keepaspectratio]{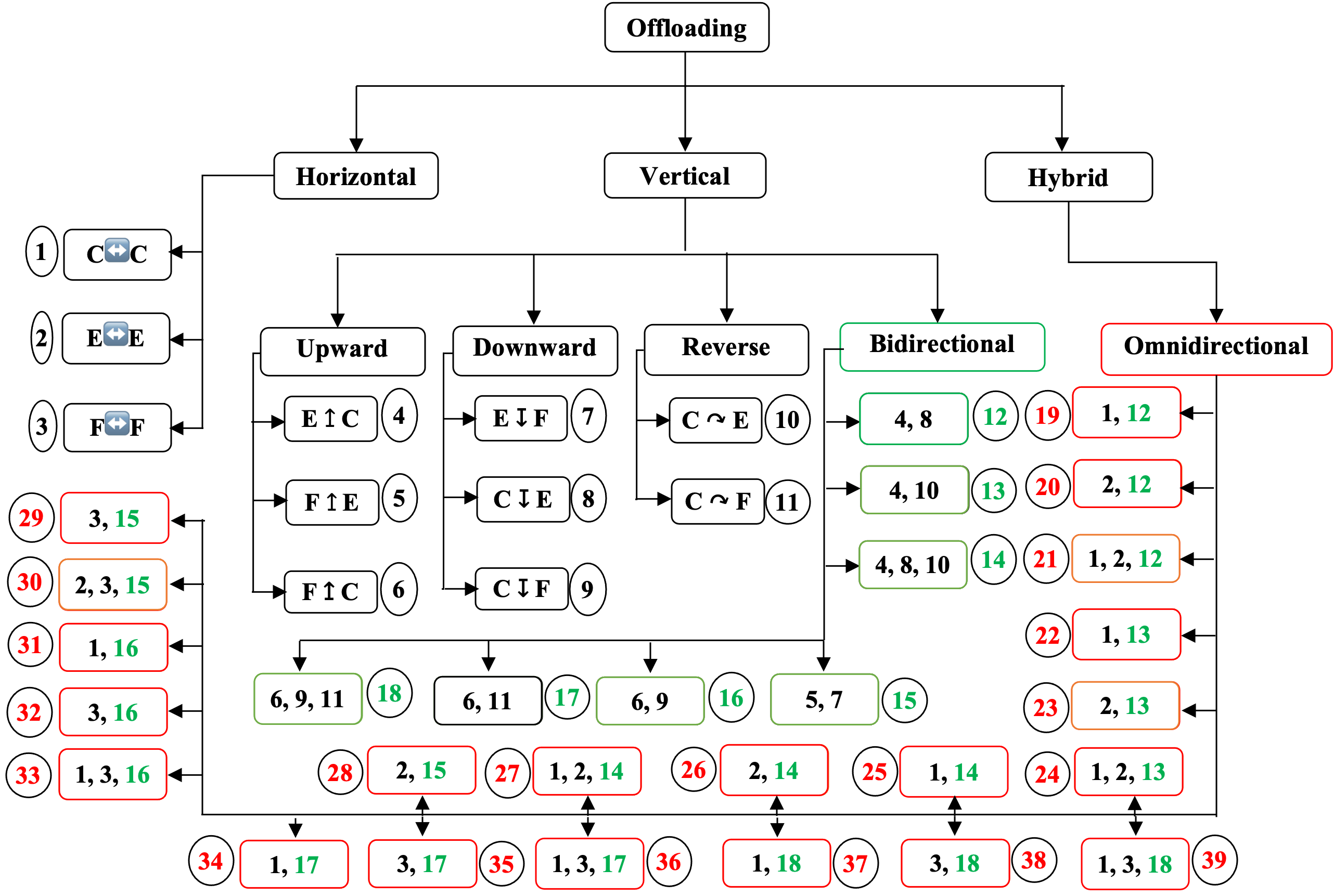}}
\caption{Classification of offloading based on federation}
\label{fig:offloading}
\end{figure*}

\subsection{Classification of Offloading}
Based on the federation agreement between entities, one entity can offload its tasks to another entity for service. This offloading can be classified into Horizontal, Vertical, or Hybrid offloading, based on different federation agreements. Our offloading classification focuses on the computation capacity and communication time perspective. However, an offloading classification can also be applied to other criteria such as storage, security, etc.

\textit{1) Horizontal Offloading.}
Horizontal offloading always takes place between two entities in the same tier with a horizontal federation agreement. As with a horizontal federation, the horizontal offloading also comes in three types, shown as \#1 to \#3 in Figure~\ref{fig:offloading}. 

\begin{enumerate}
\item In cloud-to-cloud $(C\leftrightarrow C)$ horizontal offloading, two federated clouds can offload to each other \cite{A10}. Google can offload to Amazon or vice versa. 
\item In edge-to-edge $(E\leftrightarrow E)$ horizontal offloading, two service provider in edge tiers can offload to each other \cite{B31}.  
\item In fog-to-fog $(F\leftrightarrow F)$ horizontal offloading, two computing resources in two different fogs can offload to each other \cite{A11}
\end{enumerate}

\textit{2) Vertical Offloading.}
Vertical offloading always takes place between two entities in different tiers, as for example, edge-to-cloud. There are fifteen different vertical offloading combinations from \#4 to \#18 in Figure~\ref{fig:offloading}, which can be classified into four different categories: upward (\#4 to \#6), downward (\#7 to \#9), reverse (\#7 to \#9), and bi-directional (\#10 and \#11).   

\begin{enumerate}
\item Vertical offloading occurs upward from the lower to the higher tier, which is more centralized, covers a bigger area, and has a greater computing capacity than the lower tier \cite{B23}. The possible upward offloading scenarios are edge-to-cloud $(E\mapsup C)$ fog-to-edge $(F\mapsup E)$, and fog-to-cloud $(F\mapsup C)$ offloading. 

\item When an upper tier offloads its task to a lower layer entity that is closer to the user and has lower network latency than the upper tier, it is known as downward vertical offloading. \cite{B25, E2}. The possible downward offloading scenarios are cloud-to-edge $(C\mapsdown E)$, cloud-to-fog $(C\mapsdown F)$, and edge-to-fog $(E\mapsdown F)$ offloading. These scenarios are triangular, \textit{i.e.}, the user requests are given to an upper-tier entity and then offloaded to a lower-tier entity. For example, in cloud-to-edge offloading, the cloud user gives its request to the cloud then the cloud will offload the task to the edge with which it has a federation agreement. 

\item Reverse offloading is a special type of downward vertical offloading, where the distance between two entities is relatively far, and to overcome  latency and data transfer costs associated with highly time-sensitive tasks, an entity in the upper tier can reverse offload its task to a lower-tier entity \cite{B17}. These are non-triangular offloading scenarios, \textit{i.e.}, if there is a federation between two entities in two different tiers, and if a subscriber of an entity in an upper tier is closer to an entity in a lower tier, then a users’ requests are given directly to the lower tier entity, instead of to the entity in the upper tier. For example, cloud-to-edge reverse offloading: if there is a federation between cloud and edge, the subscriber to the cloud is closer than to an edge, it can directly send the request to the edge instead of to the cloud. Cloud-to-edge $(C\curvearrowright E)$ and cloud-to-fog $(C\curvearrowright F)$ are the two reverse offloading scenarios for our system. Since edge and fog are very close to each other, we do not consider the reverse offloading scenario between them.

\item Bidirectional offloading is a combination of all possible scenarios of upward with downward offloading, upward reverse offloading, or a combination of all three, \textit{i.e.}, upward with both downward and reverse offloading. For example, the offloading scenario \#12 in Figure~\ref{fig:offloading} is a combination of offloading scenario \#4 and \#8; similarly, \#14 is a combination of \#4, \#8 and \#10. All possible bidirectional vertical offloading scenarios are shown in Figure~\ref{fig:offloading}. 
\end{enumerate}

\textit{3) Omni-directional Offloading.}
Omni-directional offloading is the combination of all possible horizontal and bidirectional offloading scenarios. For example, the \#22 offloading scenario in Figure~\ref{fig:offloading} is the combination of offloading scenarios \#1 and \#13 (a combination of \#4 and \#10). There are twenty-one different omni-directional offloading scenarios from \#19 to \#39 as shown in Figure~\ref{fig:offloading}. 

However, these scenarios are only limited to two-tier architectures. They can be further extended to three-tier  architecture by combining two two-tier architectures. To the best of our knowledge, such classification of the offloading scenarios has not previously been considered and is here set out.

\begin{table*}[!t]
\caption{{Saturation Level of Federated Systems and Different Offloading Scenarios} }
\label{table1}
\def\arraystretch{1.25}
\ignorespaces 
\centering 
\begin{tabulary}{\linewidth}{|
p{\dimexpr.1500\linewidth-2\tabcolsep}|
p{\dimexpr.2000\linewidth-2\tabcolsep}|
p{\dimexpr.35000\linewidth-2\tabcolsep}|}

\hline 
\textbf{Satutation Level} & \textbf{Federation Types} & \textbf{Offloading Types} \\
\hline 
Saturated & $C-C$ & $C\leftrightarrow C$ \\ 
\hline
Semi-saturated & $C-F$, $E-F$, $E-E$ & $E\mapsup C$, $F\mapsup C$, $F\mapsup E$, $E\leftrightarrow E$ \\ 
\hline
New & $F-F$ &  $F\leftrightarrow F$, $C\mapsdown E$, $C\mapsdown F$, $E\mapsdown F$, $C\curvearrowright E$, $C\curvearrowright F$ \\ 
\hline 
\end{tabulary}\par 
\end{table*}



\subsection{Current Research Status of Federated Architectures and Offloading}


Before doing the survey, we consider the current status of different federated architectures and offloading scenarios, which are divided into three categories, as shown in Table~\ref{table1}.
Figure~\ref{fig:offloading} shows 39 different offloading scenarios. However, out of these scenarios, 11 are core offloading scenarios that are considered for this categorization based on a one-to-one federation and offloading. Some scenarios have been addressed in many papers, which we consider a saturated scenario—for example, a  $C-C$ federation. 
In \cite{B18} Mashayekhy et al. proposed a game-theoretical model to reshape the business structure between cloud providers, which could improve their dynamic resource scaling capabilities by establishing cooperation with the federation method. They proposed a cloud federation mechanism to maximize the profit of cloud providers by reducing the utilization of computing resources. Hassan et al. \cite{B19} presented a capacity-sharing mechanism using game theory in a federated cloud environment. This mechanism may lead to a global energy sustainability policy for federated systems and can encourage such systems to cooperate. The main goal of the paper is to minimize the overall energy cost by means of a capacity sharing technique that will promote the long-term individual profit of cloud providers. 

The integration of vertical and horizontal cloud federations is discussed in \cite{B20}. In this integration, private clouds are known as secondary clouds, and are federated with each other horizontally, which become federated vertically with the public clouds, termed primary clouds. The objective of \cite{B20} is to establish stable cooperative partnerships for the federation so as to improve efficiency.  
In \cite{B22}, a distributed resource allocation problem is discussed in a horizontally dynamic cloud federation (HDCF) platform. These authors used a game theoretical solution to address this problem, to ensure mutual benefits to encourage cloud providers (CPs) to form an HDCF platform. 

Similarly, cloud-to-cloud offloading is very rare as the clouds lack capacity, capability, etc. One cloud may not have something that another one can cover, and it is then considered saturated. There are some federation architectures and offloading scenarios which have been addressed by some researchers but there is still much to address. These scenarios are termed as semi-saturated; rest are called new scenarios, in which hardly any research has been done. These three categories are shown in Table~\ref{table1}. Note that the fog as used in this paper includes any static or dynamic fog, including vehicular-fog that may have mobility. 

\begin{figure*}[!t]
\centerline{\includegraphics[width=16cm,height=13cm,keepaspectratio]{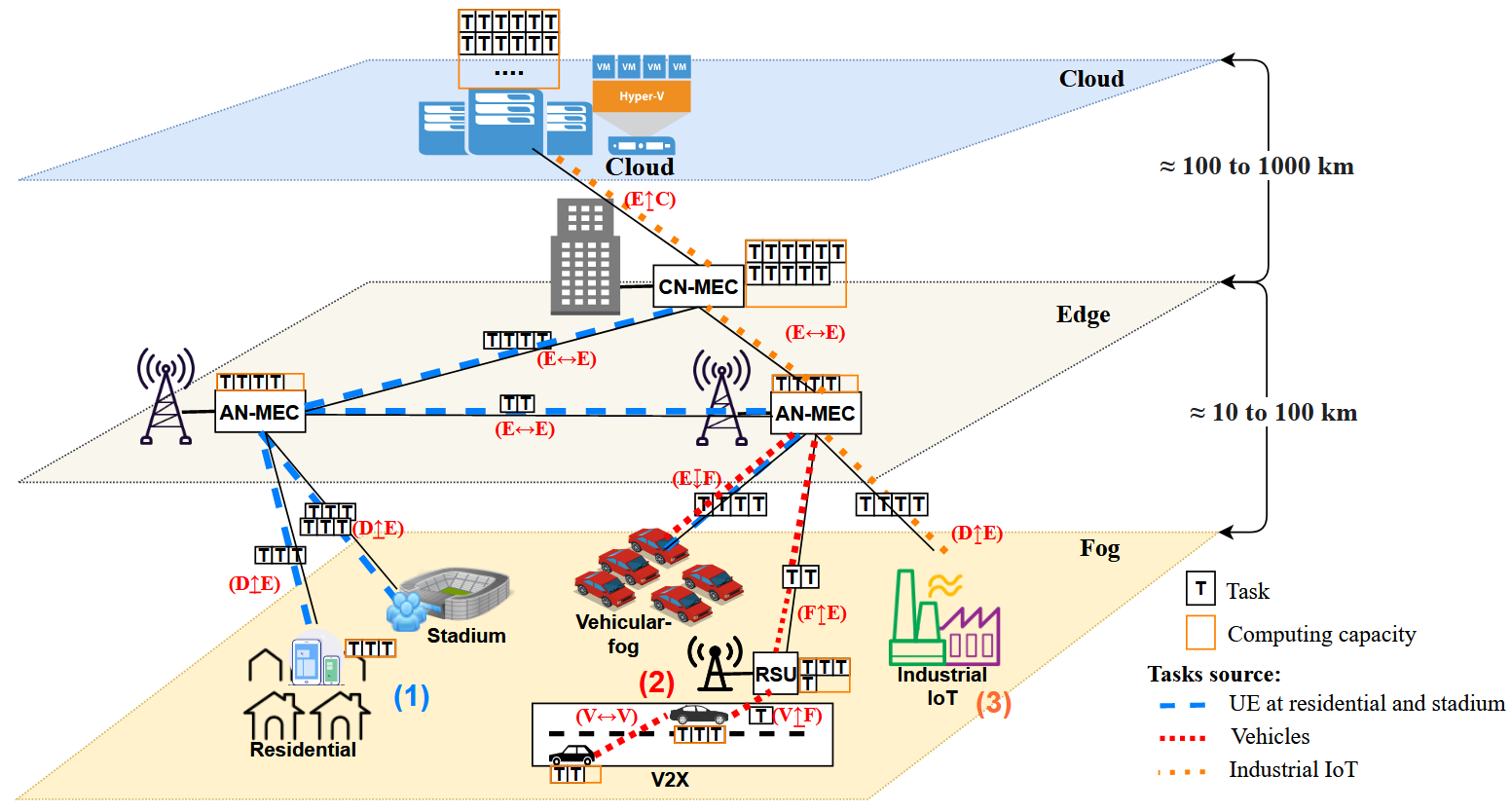}}
\caption{Task offloading in the fog-edge-cloud federation}
\label{fig:offloading_ilustration}
\end{figure*}

Figure \ref{fig:offloading_ilustration} illustrates offloading in a three-tier fog-edge-cloud federation. The fog system is comprised of a variety of devices, such as smartphones, laptops, automobiles, and road-side units (RSUs), all of which interact with one another and can even collaborate on some tasks. Between fog and cloud lies a two-tier MEC system with computing capacity behind the base stations (AN-MEC) and in a central office with core network functions (CN-MEC). Cloud computing is the top tier, with massive computing capacity but is geographically remote from UEs or data sources.

Figure \ref{fig:offloading_ilustration} shows three different offloading scenarios based on task sources. The first scenario (1) involves a heavy task or hotspot traffic at a stadium that is hosting a sporting event or music concert. The task will be offloaded from the UEs to the nearest AN-MEC. Because of the AN- MEC’s limited computational capabilities, the task can be offloaded to a less loaded AN-MEC or CNMEC, and computing delay can thus be minimized. In the second scenario (2) the vehicle generates tasks from its sensors or multimedia applications for safety and comfort. Some vehicle tasks are latency-sensitive that are part of the navigation, autopilot, accident, or alert systems. A nearby server must serve those kinds of tasks with low propagation and computing latencies . The tasks can be offloaded either horizontally to other vehicles or vertically to an RSU. If the RSU is overloaded, it will vertically offload the tasks to an AN-MEC, and the overloaded AN-MEC can offload the tasks downward to vehicular fog. The third scenario (3) describes the traffic generated by industrial IoT sensors, with some operations requiring low latency, such as robotic process automation, danger alerts, and suspicious activity alerts, and can be vertically offloaded to AN-MEC. Very large amounts of sensor data from industrial IoT can be offloaded to a cloud for future analysis.

\section{Survey on Offloading}
This section provides a summary of literature that deals with the federated environment with different offloading scenarios. Some papers discuss current surveys on cloud federation \cite{B15}\cite{B16} with cloud to cloud offloading, some edge federation \cite{B31}, some edge to cloud offloading [23], some edge to vehicular-fog offloading \cite{B25}, and some cloud to edge reverse offloading \cite{B17}. The major purpose of a federation is enhancing storage and processing capabilities. Many factors influence offloading strategies, such as the location \cite{B26}, energy \cite{B27}, and different optimization objectives. We classify this work on offloading into two categories, (a) traditional optimization techniques, that mostly focus on management plane decisions, and (b) machine learning techniques that focus on control plane decisions. \par

\subsection{Traditional Optimization}
Table \ref{table-2} lists the earlier research on traditional optimization-based offloading, according to the direction and destination of offloading.

\textit{1) Device-to-Device (D2D) Offloading.}
Some research papers focus on device-to-device (D2D) offloading \cite{B26}$-$\cite{D3}.
Wang et al. \cite{B26} investigated the mobility-assisted opportunistic computation offloading problem focusing on the patterns of contacts between mobile devices. They used the convex optimization method to determine the amount of computation tasks that can be offloaded from one device to another.
Pu et al. \cite{D2} proposed a device-to-device (D2D) fogging framework, where mobile users can dynamically and beneficially share computation and communication resources between themselves. The objective of D2D fogging is to achieve optimal energy conservation for executing the tasks of network-wide users. Yu et al. \cite{D3}  proposed a hybrid multicast-based task execution framework for multi-access edge computing (MEC). In this framework multiple devices can collaborate at the edge of a network for wireless distributed computing (MDC) and outcome sharing. Such a framework is socially aware of building effective D2D links with the objective of achieving an energy-efficient task assignment policy for mobile users. They used the Monte-Carlo search tree-based algorithm to achieve their objective. \par




\textit{2) Device-to-Fog (D2F), Device-to-Edge (D2E) and Device-to-Cloud (D2C) Offloading.}
Two papers, \cite{B27} and \cite{D1}, focused on device-to-cloud and device-to-fog offloading, respectively, while device-to-edge offloading was discussed in \cite{ADT1,ADT2,D9,D10,B28,B29,D4}.
Barbera et al. \cite{B27} tested the feasibility of mobile computation offloading in real-life scenarios. They considered an architecture where each real device is associated with a software clone on the cloud. Huang et al. \cite{ADT1} proposed a dynamic offloading algorithm based on the Lyapunov optimization that maximizes energy efficiency while preserving the required latency with face recognition applications. Zhang et al. \cite{ADT2} investigated the trade-off between energy consumption and latency for an MEC system with energy harvesting technology. They formulated the weighted sum of energy consumption and computation latency minimization of mobile device with the stability of queues and battery level, and used the Lyapunov function to ensure system stability.

Zhao et al. \cite{D9} proposed a multi-mobile-user MEC system, where multiple smart mobile devices (SMDs) can offload their tasks to an MEC server, with the objective of minimizing the energy consumption of SMDs. To optimize this, they coordinated the offloading selection, radio resource allocation, and computational resource allocation, and use the branch and bound method to solve the optimization problem.  
Wang et al. \cite{D10} investigated partial computation offloading with dynamic voltage scaling (DVS) technology in mobile edge computing, where devices can partially offload their tasks. They formulated an optimization problem with two objectives: energy consumption of SMD minimization (ECM) and latency minimization of application execution (LM). They proposed two optimal algorithms named Energy Optimal Partial Computation Offloading (EPCO), and Latency Optimal Partial Computation Offloading (LPCO) to solve the ECM, and LM problems, respectively.

To achieve minimum average delay, Liu et al. \cite{B28} adopted the Markov decision model for computational task scheduling. They proposed a searching algorithm to determine optimal scheduling. Such task scheduling is unique as the computation tasks are scheduled based on the queueing state of the task buffer, the execution state of the local processing unit, and the state of the transmission unit.
Mao et al. \cite{B29} developed a Lyapunov Optimization-based Dynamic Computation Offloading (LODCO) algorithm to minimize the execution delay and addressed task failure as the performance metric. This algorithm determines the offloading decision, the CPU-cycle frequencies for mobile execution, and the transmission power for computation offloading. However, without requiring distribution information such as computation task requests, wireless channel, energy harvesting (EH) processes, etc., these decisions depend only on the system’s current state.

Chen et al. \cite{D4} formulated a multi-user, multi-task computation offloading problem for green Mobile Edge Cloud Computing (MECC) and used the Lyaponuv Optimization approach to determine an energy harvesting policy. This policy determines how much energy is harvested from each wireless device (WD)n the task offloading schedule -- the set of computation offloading requests that can be admitted into the mobile edge cloud, the set of WDs that can be assigned to each accepted offloading request, and the amount of workload that can be processed at the assigned WDs.
In \cite{D1}, Hasan et al. present the Aura architecture, 
a highly localized and mobile ad-hoc cloud computing model using IoT devices present in the ubiquitous environment for task offloading schemes and enhancing  applications. They implemented the Aura on the Contiki platform and a simplified Map-Reduce port, which demonstrates such architecture’s feasibility.  \par

\textit{3) Device-Fog-Cloud  and Device-Edge-Cloud Vertical Upward Offloading.}
The offloading scenarios adopted in papers \cite{D8,B24,B23, ADT3, B21, ADT6, ADT7} were vertical upward, which included from device to any entity offloading and one entity to another entity offloading. Gou et al. \cite{D8} present an architecture for collaborative computation offloading over FiWi networks. They addressed the problem of cloud-MEC collaborative computation offloading to minimize the energy consumption of all the MDs while satisfying the computation execution time constraint. They proposed a distributed collaborative computation offloading scheme by adopting  game theory and analyzing the Nash equilibrium.

Sun et al. \cite{B24} addressed the latency-aware workload offloading (LEAD) problem, where they formulated a task offloading problem to minimize the average response time for mobile users. They designed the LEAD strategy and offloaded the workloads to suitable cloudlets to reduce average response times.
Tong et al. \cite{B23} proposed a hierarchical edge cloud architecture to improve the performance of mobile computing by leveraging cloud computing and offloading mobile workloads for remote execution at the cloud. For the efficient utilization of resources and workload placement, they used simulated annealing (SA) \cite{F1} to determine which programs are placed on which edge cloud servers and how much computational capacity is available to execute that program. They implemented the proposed architecture in small-scale and conducted a simulation experiment over a larger topology and evaluated the performance of a proposed workload placement algorithm.

Rodrigues et al. \cite{ADT3} proposed a heuristic offloading algorithm to determine whether tasks should be processed locally, in nearby cloudlets, or in a remote cloud, which would initially be determined by an UE.  The UE would then choose a different location to process the task and calculate the latency difference between the new location and the previous location. Each UE could make this distinction and leverage the chosen location when bidding for an offloading decision. The offloading decision with the highest bid is then chosen.

Chekired et al. \cite{B21} introduced a new scheduling model for the industrial Internet of things (IIoT) data processing, and proposed a two-tier cloud-fog architecture for IIoT applications by deploying multiple servers at the fog tier. The objective of this architecture was to minimizing communication and data processing delays in IIoT systems.
Resource allocation and offloading optimization for heterogeneous real-time tasks were carried out by means of an adaptive queueing weight (AQW) resource allocation policy in \cite{ADT6}. A trade-off between throughput and task completion ratio optimization was also achieved by taking laxity and completion times into account when designing the offloading policy.
Adhikari et al. \cite{ADT7} designed a novel delay-dependent Priority-Aware Task Offloading (DPTO) algorithm for scheduling and handling IoT device tasks in an appropriate computing server. The computing locations were chosen based on the types of task deadlines, which were classified as soft and hard-deadline tasks.

\textit{4) Device-Edge-Cloud Hybrid Offloading.}
Hybrid offloading was discussed in \cite{B30, ADT4, B32} which included device-edge vertical offloading. 
Tran and Pompili \cite{B30} formulated a mathematical model for the joint optimization of task offloading and resource allocation in MEC. In this work they did not only account for the allocation of computing resources but also for the allocation of the transmission power of mobile users.

\begin{figure*}[!t]
\centerline{\includegraphics[width=18cm,height=16cm,keepaspectratio]{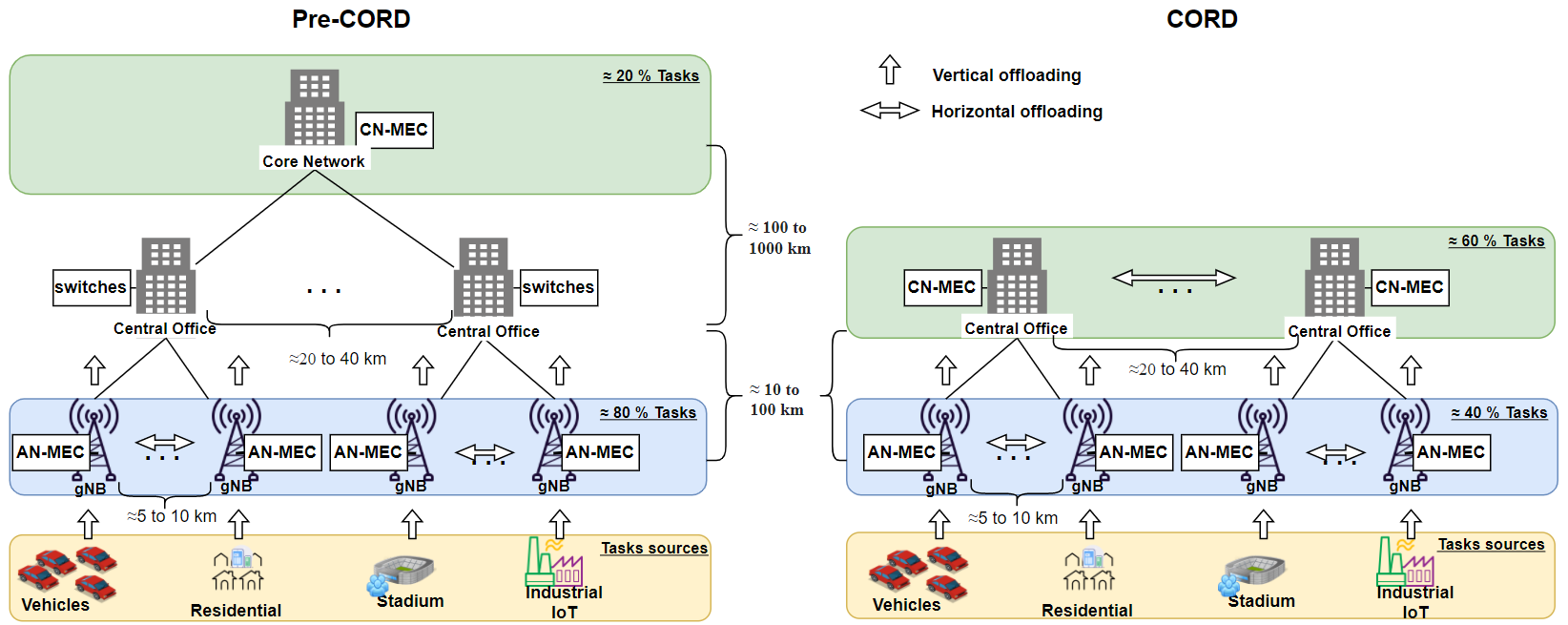}}
\caption{Ratio-based offloading: pre-CORD vs. CORD}
\label{fig:offloadingex}
\end{figure*}

The two-tier MEC architecture proposed by yahya et al. \cite{ADT4} is comprised of an access network (AN) and a core network (CN) MEC. CN-MEC has greater capacity but is less wide spread than AN-MEC. Two-phase optimization was used to achieve capacity optimization by modifying the offloading ratio and capacity iteratively. For hot-spot traffic, offloading and scaling were merged into short-term and long-term solutions. They considered both vertical, device-edge, and horizontal offloading between edges. In a comparison between pre-CORD and CORD, shown in Figure \ref{fig:offloadingex}, a trade-off between computing and communication latency was introduced for different distances of the CN MEC which affected the task processing distribution.  
Thai et al. \cite{B32} proposed workload and capacity optimization to minimize computation and communication costs for cloud-edge federated systems, by taking consideration vertical and horizontal offloading. They designed a branch and bound algorithm with parallel multi-start search points to solve this problem.\par

\begin{figure}[!t]
\centerline{\includegraphics[width=9 cm,height=8cm,keepaspectratio]{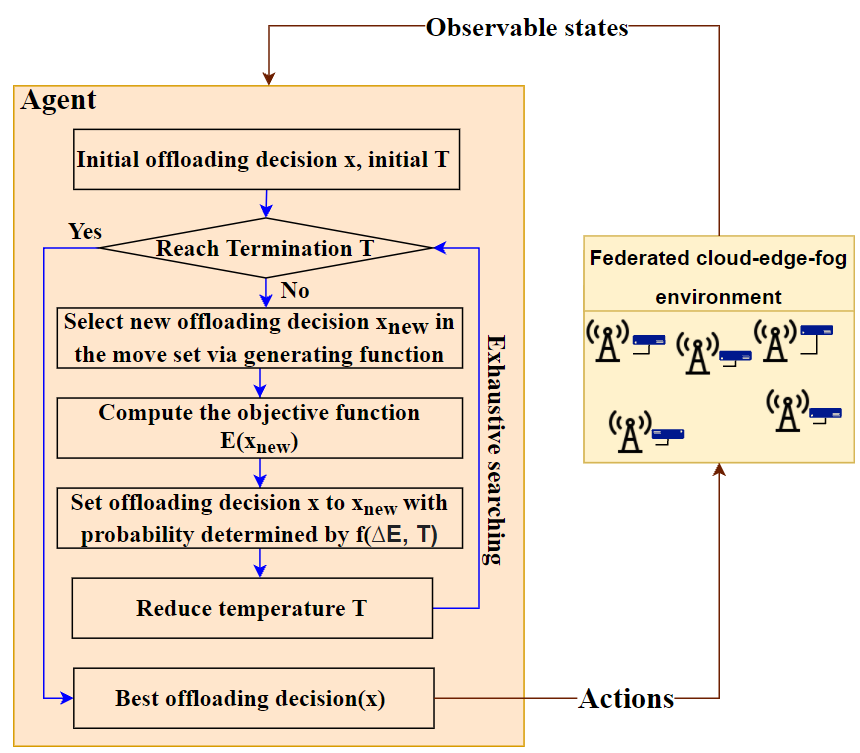}}
\caption{Simulated Annealing-based offloading illustration}
\label{fig:traditional_ilustration}
\end{figure}

Villar et al. \cite{B17} introduced osmotic computing, a new paradigm for edge and cloud integration. In their research they develop the concept of reverse offloading, where not only can an edge  offload its tasks to the cloud, but the cloud can also reverse offload time-sensitive tasks to edges. 
A two-tier cloud-edge federated architecture was proposed by Kar et al. \cite{E1}, who considered edge-to-edge horizontal offloading and edge-to-cloud vertical offloading together with cloud-to-edge reverse offloading. They formulated an optimization problem with the objective of minimizing costs where latency was the key constraint, and used simulated annealing to solve it. As shown in Figure \ref{fig:traditional_ilustration}, the simulated annealing technique gathers system information and carries out an exhaustive search into acquiring the best offloading decision. \par

\textit{5) Fog-Edge-Cloud Vertical Upward Offloading.}
Some papers \cite{ADT5, B34}, \cite{D6}-\cite{D5} focus on entity to entity upward offloading, and some adopt hybrid offloading scenarios \cite{D7}-\cite{B31}. Fantacci and Picano \cite{ADT5} carried out queueing analysis of cloud-fog-edge computing infrastructure and proposed a heuristic to determine offloading ratios and computing capacities at fog, edge, and cloud. Kar et al. \cite{BK2021} considered a federated architecture with mobile device, edge, cloud, and vehicular-fog together. They used the queueing theory to analyze the performance to minimize QoS violation probability, and used a subgradient searching algorithm to determine the optimal probabilities.

An intelligent offloading method (IOM) for smart cities, conserving privacy, improving offloading efficiency, and promoting edge utility, was proposed to address privacy disclosure in Xu et al. \cite{B34}. 
These authors used the ant colony optimization (ACO)  \cite{F2} method to achieving the trade-offs between minimizing service response time, energy optimization and maintaining load balance while ensuring privacy preservation during service offloading. An energy-efficient computation offloading mechanism for MEC in 5G heterogeneous networks was proposed in \cite{D6}. They formulated the energy minimization problem of an offloading system, where both task computing and file transmission energy costs were considered. 

Lu et al. \cite{D5} addressed the problem of computation offloading by using edge computing. They formulated the problem as a two-stage Stackelberg game problem and show that it achieves a Nash equilibrium. Their objective was to maximize cloud service operators’ and edge server owners’ utilities by obtaining optimal payment and computation offloading strategies with low delay. 
Ma et al. \cite{B33} proposed a cloud-assisted framework in MEC, termed Cloud Assisted Mobile Edge computing (CAME), to minimize resource costs by combing queueing network and convex optimization theories. They solved the convex problem by using Karush-Kuhn-Tucker (KKT) conditions, and augmented reality to represent delay-sensitive and computation-intensive mobile applications.

Jiao et al. \cite{D7} presented an integrated framework for computation offloading and resource allocation in MEC networks, where both single and multi-cell networks were taken into consideration. To minimize energy consumption and delay, they proposed an energy-aware offloading scheme that considers both computation and communication resource allocation.  
In \cite{B31} a horizontal edge federation was proposed together with UE to edge and edge to cloud vertical offloading scenarios. They experimentally showed that an edge federation model improves the quality of experience (QoE) of end-users and saves on the costs of edge infrastructure providers (EIPs). \par

\textit{6) Vehicular-Fog and V2X Offloading.}
The single edge to vehicular-fog task offloading problem was addressed in \cite{B25} where an iterative greedy algorithm was used to solve the optimization problem. 
Yen et al. \cite{E2} proposed a decentralized offloading configuration protocol (DOCP) for single edge to vehicular-fog offloading, with a matching protocol between multiple edge systems to resolve resource contention when resources from the same vehicular-fog were requested simultaneously.

Offloading optimization for vehicular-to-everything (V2X) systems was addressed in \cite{ADT8, ADT9}. Zhang et al. \cite{ADT8} considered hybrid offloading between vehicles and fogs, and formulated a mixed-integer, nonlinear programming (MINLP) solution for optimizing both user association and radio resource allocation in vehicular networks (VNET). To obtain a globally optimal solution, this MINLP problem was transformed by applying norm theory to non-convex nonlinear fraction optimization and then showed to be equivalent to convex optimization using weighted minimum mean square error (WMMSE) and Perron-Frobenius theory. Wang et al. \cite{ADT9} proposed a real-time traffic management algorithm for fog-based Internet-of-Vehicle (IoV) systems. This consisted of a three-tier architecture of fog, cloudlet, and cloud for providing computing resources to traffic management systems. They also looked into vertical offloading optimization between fog, cloudlet, and cloud.


\begin{table*}[!t]
\renewcommand{\arraystretch}{1.25}
\centering
\caption{Analysis of Recent Research on Offloading in the Federated Systems with Traditional Optimization}
\label{table-2}
\begin{tabular}{|l|l|l|l|l|l|l|l|l|l|}
\hline
 &  & \multicolumn{4}{l|}{\textbf{Metrics}} &  &  &  & \\ \cline{3-6} 
\textbf{\multirow{2}{*}{References}}  & \textbf{\multirow{2}{*}{Offloading Types}} & \begin{turn}{-90} \textbf{Cost} \end{turn}& \begin{turn}{-90} \textbf{Energy} \end{turn} & \begin{turn}{-90} \textbf{Capacity\;} \end{turn} & \begin{turn}{-90} \textbf{Latency} \end{turn} & \textbf{\multirow{2}{*}{Approach}} & \textbf{\multirow{2}{*}{Method}} & \textbf{\multirow{2}{*}{Evaluation}} &  \textbf{\multirow{2}{*}{Application}}\\ \hline

\cite{B26} & \multirow{3}{*}{$D2D$} &  & $\checkmark$ &  & & Exact & Convex optimization & Simulation & Offloading in realistic \\ 
&&&&&&& &  & human mobility scenario  \\ \cline{1-1} \cline{3-10}
\cite{D2} &&& $\checkmark$ & $\checkmark$ & & Analysis & Lyapunov optimization &Simulation & Fogging framework \\\cline{1-1} \cline{3-10}
\cite{D3} & & & $\checkmark$ & $\checkmark$ &  & Scheme  & Tree search algorithm & Simulation & Face recognition\\ \hline

\cite{B27} & $D\mapsup C$ &  & $\checkmark$ &  &  & No approach  & Real test-bed & Experimental & MCC application \\  \hline

\cite{ADT1} &  \multirow{7}{*}{$D\mapsup E$} &  & $\checkmark$ &  & $\checkmark$ & Analysis  & Lyapunov optimization & Simulation & Face recognition\\ \cline{1-1} \cline{3-10}

\cite{ADT2} &  &  & $\checkmark$ &  & $\checkmark$ & Heuristic
 & ODLOO & Simulation & Generic user applications\\ \cline{1-1} \cline{3-10}

\cite{D9} &  &  & $\checkmark$ &  & $\checkmark$ & Analysis  & Branch and bound & Simulation & Smart mobile device (SMD)\\ \cline{1-1} \cline{3-10}
\cite{D10} &  &  & $\checkmark$ &  & $\checkmark$ & Analysis  & EPCO algorithm, &Simulation& Data partitioned in SMD \\ 
&&&&&&&  LPCO algorithm  & & \\ \cline{1-1} \cline{3-10}

\cite{B28} &  & & $\checkmark$  & & $\checkmark$  & Policy  & One-dimensional & Simulation & MEC systems\\ 
&&&&&&&  search algorithm  & & \\ \cline{1-1} \cline{3-10}

\cite{B29} &  &  & $\checkmark$  &  & $\checkmark$ & Analysis  & Lyapunov optimization & Simulation & Energy harvesting for \\ 
&&&&&&& && devices \\ \cline{1-1} \cline{3-10}

\cite{D4} &  &  & $\checkmark$ & $\checkmark$ &  & Policy  & Lyaponuv optimization & Simulation & Multi-user multi-tasking \\ \hline
\cite{D1} & $D\mapsup F$ & & $\checkmark$ & $\checkmark$ & & Scheme  & Aura architecture & Experimental & Prototype design\\ \hline

\cite{B34} & \multirow{2}{*}{$F\mapsup E$} &  & $\checkmark$  & & $\checkmark$  & Analysis  & Ant colony optimization & Simulation & Smart city\\ \cline{1-1} \cline{3-10}

\cite{D6} & &  & $\checkmark$ & & $\checkmark$ & Scheme & EECO scheme & Simulation &  5G heterogeneous networks \\ \hline

\cite{D5} & \multirow{2}{*}{$E\mapsup C$} & $\checkmark$ & & $\checkmark$ & & Analysis   & Game theory & Simulation & Payment strategy in edge  \\
&&&&&&&&&computing\\ \cline{1-1} \cline{3-10}
\cite{B33} &  & $\checkmark$ &  & $\checkmark$ & & Analysis & KKT conditions &Simulation & Augmented reality\\ \hline

\cite{B25} & \multirow{2}{*}{$E\mapsdown F$} & $\checkmark$ &  &  & $\checkmark$ & Heuristic  & Iterative greedy & Simulation & Intelligent transportation  \\ 
&&&&&&& & & systems \\ \cline{1-1} \cline{3-10}
\cite{E2} & & $\checkmark$ &  &  & $\checkmark$ & Heuristic  & Iterative greedy, & Simulation & Intelligent transportation  \\ 
&&&&&&&  DOCP & & systems \\ \hline

\cite{D8} & \multirow{3}{*}{$D\mapsup E$, $E\mapsup C$} &  & $\checkmark$ & $\checkmark$ & $\checkmark$ & Scheme  & Game theory & Simulation & FiWi networks\\ \cline{1-1} \cline{3-10}
\cite{B24} &  &  &  & $\checkmark$ & $\checkmark$ & No approach  & LEAD algorithm & Simulation & MCC application\\ \cline{1-1} \cline{3-10}
\cite{B23} & &  &  & $\checkmark$ & $\checkmark$ & Analysis   & Simulated annealing & Experimental, & Traffic engineering \\  \cline{1-1} \cline{3-10}

\cite{ADT3} & &  &  &  & $\checkmark$ & Heuristic  & Iterative search algorithm & Simulation & Generic user applications \\ \hline

\cite{B21} & \multirow{2}{*}{$D\mapsup F$, $F\mapsup C$} &  &  & $\checkmark$ & $\checkmark$ & Analysis  & Simulated annealing & Simulation & Industrial IoT \\  \cline{1-1} \cline{3-10}

\cite{ADT6} & &  &  & $\checkmark$ & $\checkmark$ & Heuristic   & LTS-AQW & Simulation & Real-time applications \\   \cline{1-1} \cline{3-10}

\cite{ADT7} & &  &  &  & $\checkmark$ & Heuristic   & DPTO & Simulation & IoT applications \\  \hline

\cite{ADT5} & $D\mapsup F$, $F\mapsup E$ &  &  &  & $\checkmark$ & Analysis  &  Iterative greedy  &Simulation & Generic user applications \\ 
& $E\mapsup C$ &&&&&& && \\ \hline

\cite{BK2021} & $D\mapsup F$, $D\mapsup E$ &  &  & $\checkmark$ & $\checkmark$ & Analysis  &  Subgradient iterative   &Simulation & MCC applications and ITS\\ 
& $E\mapsup C$ &&&&&& method && \\ \hline

\cite{D7} & $F\mapsup E$, $E\mapsup C$ &  & $\checkmark$ &  & $\checkmark$ & Scheme  &  Iterative search algorithm & Simulation & Multi-cell MEC networks\\ \hline

\cite{ADT8} & $V\leftrightarrow V$, $V\mapsup F$ &  & &  & $\checkmark$ & Heuristic  &  Iterative searching & Simulation & Multimedia applications\\ \hline

\cite{ADT9} & $V\mapsup F$, $F\mapsup E$ &   &  &  & $\checkmark$ & Analysis  & Branch-and-bound and & Simulation & Traffic management system\\ 
&$E\mapsup C$ &&&&&&  Edmonds–Karp  & & \\ \hline 

\cite{B30} & \multirow{2}{*}{$D\mapsup E$, $E\leftrightarrow E$}& & $\checkmark$ & $\checkmark$ & $\checkmark$  &  Heuristic  & Bisection method &Simulation & Multi-cell  wireless network\\ \cline{1-1} \cline{3-10}
\cite{ADT4} & &  & $\checkmark$ &  & $\checkmark$ & Heuristic   & Two-phases iterative  & Simulation & URLLC, eMBB, and MMTC \\  
&&&&&&&  optimization  & & \\ \hline

\cite{B32} & $D\mapsup E$, $E\leftrightarrow E$ & $\checkmark$ &  & $\checkmark$ & $\checkmark$ & Analysis  &  Branch and bound  &Simulation & Traffic engineering \\ 
& $E\mapsup C$ &&&&&& && \\ \hline

\cite{B31} & $E\mapsup C$, $E\leftrightarrow E$ & $\checkmark$ &  & $\checkmark$ &  & Scheme  & Dynamic algorithm &Experimental & Traffic engineering\\ \hline

\cite{B17} & $E\mapsup C$, $C\curvearrowright E$ &  &  & $\checkmark$ & $\checkmark$ & No approach & Architecture &  & Osmotic computing\\ \hline
\cite{E1} & $E\leftrightarrow E$, $E\mapsup C$ & $\checkmark$  &  & $\checkmark$ & $\checkmark$ & Analysis  & Simulated annealing & Simulation & Traffic engineering\\ 
& $C\mapsdown E$, $C\curvearrowright E$ &&&&&& && \\ \hline


\end{tabular}

\begin{tablenotes}
      \small
	\item $D$: Device; $D2D$: Device to Device; MCC: Mobile Cloud Computing; MEC: Multi-access Edge Computing; IoT: ; EPCO: Energy-Optimal Partial Computation Offloading; LPCO: Latency-Optimal Partial Computation Offloading; LEAD: Latency-Aware workloAd offloaDing; EECO: Energy-Efficient Computing Offloading; KKT: Karush-Kuhn-Tucker; DOCP: Decentralized Offloading Configuration Protocol
\end{tablenotes}

\end{table*}

A summary of the above-discussed literature is given in Table~\ref{table-2}. The organization of the comparison table is as follows. We discussed different core offloading methods used in the papers, including device-to-device (D2D) and device to other entities. Four standard metrics, i.e., cost, energy, capacity, and latency, are considered that are commonly used in most literature. Although there are other factors such as QoS, load balance, intensive, etc., those are not presented in the table but are already addressed in descriptions. Each paper has different approaches such as exact, analysis, scheme, policy, heuristic, and evaluation method presented in the table.

\subsection{Machine Learning}

\begin{table*}[!t]
\renewcommand{\arraystretch}{1.5}
\centering
\caption{Comparisons of Machine Learning Algorithms for Offloading in Federations}
\label{tabel-4}
\resizebox{\textwidth}{!}{\begin{tabular}{|l|p{1.5cm}|p{1.25cm}|c|p{1.25cm}|c|c|c|c|p{2.25cm}|p{1.5cm}|}
\hline
\multicolumn{1}{|c|}{\multirow{2}{*}{\textbf{ML Approaches}}} & \multicolumn{1}{c|}{\multirow{2}{*}{\textbf{Paper}}} & \multirow{2}{1.25cm}{\textbf{Online learning}} & \multirow{2}{*}{\textbf{Supervisor}} & \multicolumn{1}{c|}{\multirow{2}{1.5cm}{\textbf{Learning object}}} & \multicolumn{2}{c|}{\textbf{Model   dependence}} & \multicolumn{2}{c|}{\textbf{Learning direction}} & \multicolumn{1}{c|}{\multirow{2}{2cm}{\textbf{Performance}}} & \multicolumn{1}{c|}{\multirow{2}{1.5cm}{\textbf{Adaptability}}} \\ \cline{6-9}
\multicolumn{1}{|c|}{} & \multicolumn{1}{c|}{} &  &  & \multicolumn{1}{c|}{} & \multicolumn{1}{l|}{\textbf{Model based}} & \multicolumn{1}{l|}{\textbf{Model free}} & \multicolumn{1}{l|}{\textbf{Value based}} & \multicolumn{1}{l|}{\textbf{Policy based}} & \multicolumn{1}{c|}{} & \multicolumn{1}{c|}{} \\ \hline
Supervised ML & \cite{M1,M3,M4} & \multirow{2}{*}{ } & \multirow{2}{*}{Yes} & \multirow{2}{*}{Dataset} & \multirow{2}{*}{$\checkmark$} & \multirow{2}{*}{} & \multirow{2}{*}{$\checkmark$} & \multirow{5}{*}{} & \multirow{2}{2.5cm}{Depend on data and learning algorithm} & \multirow{2}{2cm}{No} \\ \cline{1-2}
DL &\cite{M9,M13,M16,M17} &  &  &  &  &  &  &  &  &  \\ \cline{1-8} \cline{10-11} 
MAB &\cite{M20} & \multirow{5}{*}{     $\checkmark$} & \multirow{5}{*}{No} & \multirow{5}{*}{Environment} & \multirow{5}{*}{} & \multirow{5}{*}{$\checkmark$} & \multirow{3}{*}{$\checkmark$} &  & \multirow{5}{2cm}{Depend on the experience} & \multirow{5}{2cm}{Through exploration} \\ \cline{1-2}
(RL)Q-Learning &\cite{M8, M12} &  &  &  &  &  &  &  &  &  \\ \cline{1-2}
(DRL)DQN &\cite{M5,M11,M15,M18, M26} &  &  &  &  &  &  &  &  &  \\ \cline{1-2} \cline{8-9}
(DRL)E2D &\cite{M6} &  &  &  &  &  & \multicolumn{2}{c|}{\multirow{2}{*}{$\checkmark$}} &  &  \\ \cline{1-2}
(DRL)DDPG &\cite{M2,M4,M7,M10,M19, ADM2} &  &  &  &  &  & \multicolumn{2}{c|}{} &  &  \\ \hline
\end{tabular}}
\end{table*}

In federation architecture, an offloading module which distributes tasks from one entity to other entities or tiers is part of the control plane. The decision of task offloading in an extensive federated system must be carried out quickly, usually in seconds. Traditional optimization, such as a non-convex algorithm, carries out an exhaustive search that takes a long time to converge and violates the delay requirements of tasks \cite{M22}. Furthermore, a traditional optimization algorithm needs complete system information to determine offloading, which some federations may not provide. Intensive system monitoring that provides complete information for determining offloading action in a federation is not trivial because each provider uses different devices, protocols, and operating systems. Some applications provided by federation may also have different requirements \cite{M1, M2, M4, M10}. Machine learning is a suitable approach to address such offloading problems in a highly dynamic system with some unknown information.

Machine learning- based (ML) offloading can automatically improve its actions by learning from the collected data (dataset) or interacting with the environment. Some ML approaches are compared in Table \ref{tabel-4}. Supervised ML and Deep Learning (DL) update their model’s weight in order to execute the best offloading decision by learning from previous data, which is categorized as offline learning. A well-labelled data-set has first to be constructed before being provided to the ML algorithms. Some providers may restrict the details of their data-sets because of security. Another way to train an offloading model is through online interaction between a learning agent and the environment, which is termed Reinforcement Learning (RL). The learning agent observes an environment’s conditions to determine an offloading action. An environment will then give positive and negative feedback on the taken action, termed reward and punishment.  In essence, an agent memorizes this interaction in the form of a table to decide the best action to take in the future. In a large system, such as a federation, maintaining agent interactions in a table leads to a scalability problem. Deep Reinforcement Learning exchanges the table with a neural network which can predict the reward of an actions for given environment’s state.

These concepts are classified into 11 types in Table \ref{tabel-5}, depending on their offloading direction and destination in the federated fog, edge, and cloud. 

\textit{\textcolor{black}{1) Device-Edge-Cloud Offloading.}} Junior et al. \cite{M1} considered cloud capacity to provide an external computation capacity to UE applications, such as image editor, face detection, and online games. They proposed a device application architecture which consists of middle-ware, a profiler and a decision engine to determine offloading policy. Multiple classifiers were evaluated to find the most accurate one, with the objective of achieving high offloading accuracy, which would result in low latency and energy efficiency.

\begin{figure}[!t]
\centerline{\includegraphics[width=9cm,keepaspectratio]{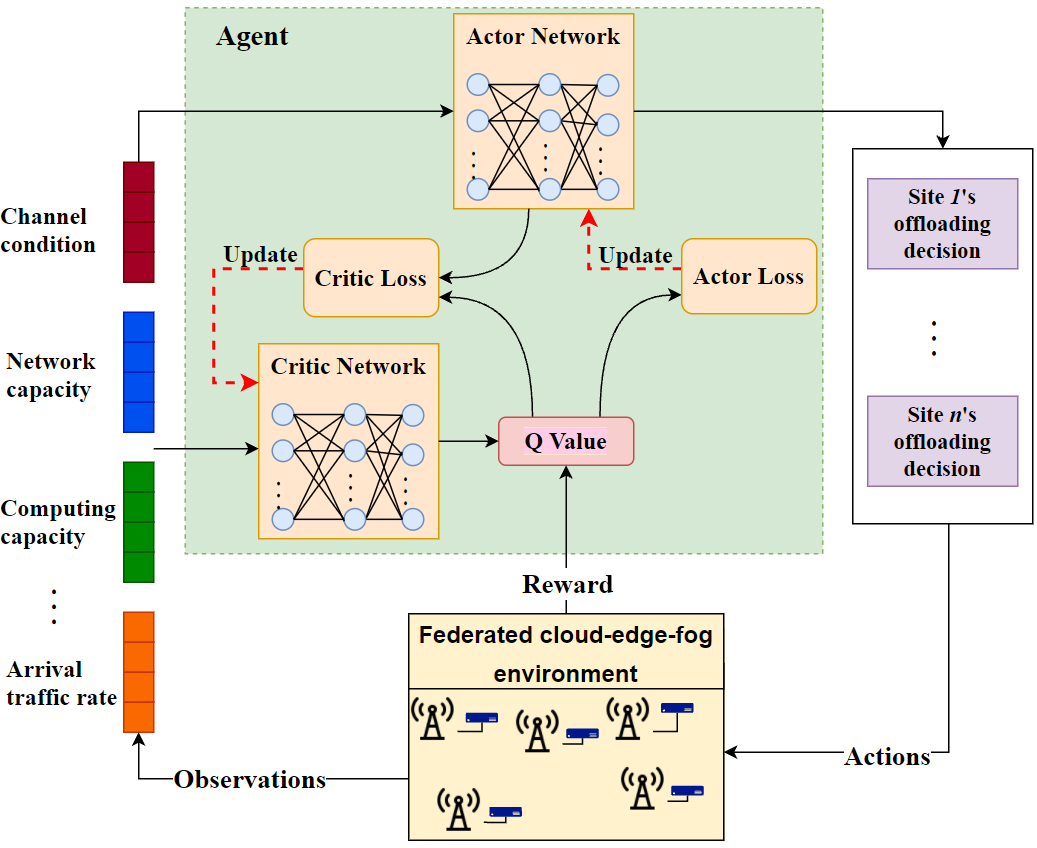}}
\caption{DRL-based offloading in the fog-edge-cloud federation}
\label{fig:RL_ilustration}
\end{figure}

Studies \cite{M2, M3, M4, M5, M6, M7, M8, M9, M10, M11, M12, M13} dealt with device to edge offloading. Other than offloading policy, Saguil and Azim \cite{M2} also considered  caching strategy to locate application codes and data. Q-learning and DQN-based algorithm solved this joint optimization problem. 

Li et al. \cite{M3} considered task deadline time in determining task offloading policy. Tey proposed an E2D DRL to derive the best offloading policy and solve the scalability problem of DQN action space. Wang et al. \cite{M4} optimized a UAV trajectory and offloading decision, which included discrete and continuous variables, by using multi-agent reinforcement learning. DDPG technique was chosen because it solved the overestimation problem of RL and works in high-dimensional action spaces. Figure \ref{fig:RL_ilustration} shows the DDPG algorithm overview. This is implemented on an agent that determines the optimal offloading decision based on federated system data such as channel status, arrival traffic information, computation and networking capacity. DDPG makes use of Actor and Critic neural networks. Actor networks predict the optimal action for a given state, whereas Critic networks predict the value of state-action pairs. The Q-value provides the discounted total future reward for the current state-action pair. By satisfying Bellman's equation, the critic network learns this value. 

Joint offloading and resources allocation optimization was carried out by Yang et al. \cite{M5}, who applied single and multi-agent reinforcement learning to optimize caching and offloading decision, and LSTM to predict task popularity in pre-processing.  Ale et al. \cite{M6} addressed the computation offloading problem of a multi-server MEC system by using DRL. Reformatting the features and storing in a tree-like data structure were carried out to accelerate DRL’s convergence time. 

DRL was used in \cite{M7} to group NOMA's UEs to minimize offloading energy, by minimizing multiple access interference. Chen et al. \cite{M8} extended DDPG with a temporal feature extraction network (TFEN) and a rank-based prioritized experience replay (rPER) to achieve training stability and reduce convergence time. Guo et al. \cite{M9} used a binary-tree based supervised ML to construct an intelligent offloading task with high accuracy and low complexity. Multichannel access problems arise in multi-user offloading when some mobile users utilise the same channel, which then  results in longer transmission latency due to interference. Cao et al. \cite{M10} used multi-agent reinforcement learning to derive the best offloading policy. The user device plays the role of an agent that observes channels condition to determine  offloading policy. Yang et al. \cite{M11} combined offline learning based on a feed-forward neural network and online inference to derive an offloading strategy in near real-time. Zhang et al. \cite{M26} enhanced the DQN algorithm with a heuristic offloading technique in order to reduce both latency and energy consumption. The objective of using an heuristic algorithm was to minimize convergence time. DQN and DDPG were compared in optimizing offloading decisions by \cite{M12}. DDPG outperform DQN in term of convergence time and performance in minimizing system latency. Li et al. \cite{M13} integrated a Lyapunov optimization with DDPG to achieve a long-term objective in online offloading.

Other than vertical offloading from UE devices to edge, He et al. \cite{M14} also considered horizontal offloading between UE devices. They applied QoE in determining offloading policy, and defined task priority assignment, redundant task elimination, and defined task scheduling to achieve optimum QoE. Since the offloading decision involved continuous action space, the DDPG-based method of DRL was used. 

The edge and cloud federation was considered in \cite{M15, ADM2, M16}. Sun et al. \cite{M15} proposed a machine learning model cooperatively trained the two-tier edge-cloud architecture. The Industrial Internet of Things (IIoT) devices was used to the determined offloading their tasks to an edge or cloud, depending on which would satisfy the tasks in terms of latency. This is categorized as upward vertical offloading. If both edge and cloud could not meet the latency requirements, the IIoT device processed the task locally. Hou et al. \cite{ADM2} applied Cybertwin to coordinate resources between end-edge-cloud. Cybertwin functions as an intelligent agent that makes the offloading decisions necessary to accomplish the objectives of maximizing processing efficiency and task completion rate. They classified IoT applications into either delay-sensitive or delay-tolerant. To maximize processing efficiency, a joint optimization of hierarchical task offloading and resource allocation based on MADDPG was proposed. The offloading agent was trained in a federated fashion. These approaches share only a trained model during the training process, avoiding the sharing of local data, which could jeopardize privacy.
Zhang et al. \cite{M16} discussed  downward vertical offloading, which was carried out by multi-cloud systems to edge servers or mobile devices. Multiple clouds compete with each other to access network and MEC resources. A distributed offloading problem arises in a system with no centralized control, such as a multi-cloud system. They also proposed multi-agent Q-learning to determine optimum offloading policy, which minimizes system latency.

\textit{\textcolor{black}{2) Device-Fog-Cloud  Offloading.}}
Devices to fog offloading was discussed in \cite{M17, M18, ADM1}. Saguil and Azim \cite{M17} considered worst-case execution time in determining offloading policy to fog node. Their objective was to minimize the execution time of time-consuming ML tasks generated by an embedded system. Li et al. \cite{M18} considered time-varying task characteristics and fog node capability in determining the offloading policy of a DQN-based algorithm. Alelaiwi et al. \cite{M19} also considered a fog and cloud federation, particularly horizontal offloading between fogs. DL was used to predict response times at multi-tier fog, edge, and cloud, which were task- offloading destinations. They applied Deep Belief Network (DBN) and logistic regression layer, which accepted processing, memory and link capacity as inputs. Ren et al. \cite{ADM1} used MADRL-based DQN to determine the best fog access point (F-AP) to serve as an IIoT node request. Because of the capacity constraints of the F-AP, some IIoT device requests have to be offloaded to the cloud, a decision made using a low-complexity greedy algorithm.


\textit{\textcolor{black}{3) V2X Offloading.}}
A federation which included a V2X system were considered in \cite{M20, M21, M22, M23, M24, M25}. The papers \cite{M20, M21, M22} optimized vertical offloading from vehicles to edge servers. Online and offline learning were used by Fan et al. \cite{M20} to maximize user and access network throughput. Pareto optimization mapped the vehicles and access points, and the optimal results were used to construct a data set for DNN model training. An online stage used the output of the trained DNN model to predict the optimal association between vehicles and access points. 

Ning et al. \cite{M21} optimized offloading decisions and resource allocations jointly in a vehicular edge system, with the objective  of maximizing QoE. DQN-based offloading task scheduling, which also considers user mobility, was proposed. Sonmez et al. \cite{M22} proposed an ML-based task orchestrator for vehicular edge systems, including LAN, MAN, WAN networks. An ML-based task orchestrator guarantees a task being served successfully (in time) and in the lowest service time, and Xie et al. \cite{M23} considered not only vertical offloading between vehicles and edge, but also considered horizontal offloading between vehicles. Vehicles, which have tasks to offload, learned the environment with the multi-armed bandit (MAB) method to  determine offloading policy, which resulted in lower average latency than the Greedy algorithm. 

The papers \cite{M24, M25} considered a fog and cloud federation to accommodate offloading tasks from vehicles. Khayyat et al. \cite{M24} used deep-Q learning, which has multiple DNN that can work in parallel to obtain the optimal offloading decision. In their environment, five DNNs would outperform a single DNN. Gao et al. \cite{M25} addressed the task dependency offloading problem by using multi-agent reinforcement learning. Their objective was minimizing energy and latency of the offloading task. LSTM was integrated into an RL to alleviate an incomplete environment’s state.

\begin{table*}[!t]
\renewcommand{\arraystretch}{1.5}
\centering
\caption{Analysis of Recent Research on Offloading in the Federated Systems with Machine Learning}
\label{tabel-5}
\resizebox{\textwidth}{!}{\begin{tabular}{|c|l|l|l|l|l|l|c|l|l|}
\hline
\multirow{2}{*}{\textbf{References}} & \multicolumn{1}{c|}{\multirow{2}{*}{\textbf{Offloading Types}}} & \multicolumn{4}{c|}{\textbf{Metrics}} & \multicolumn{1}{c|}{\multirow{2}{*}{\textbf{Method}}} & \multirow{2}{*}{\textbf{Agent}} & \multicolumn{1}{c|}{\multirow{2}{*}{\textbf{Evaluation}}} & \multicolumn{1}{c|}{\multirow{2}{*}{\textbf{Application}}} \\ \cline{3-6}
 & \multicolumn{1}{c|}{} & \begin{turn}{-90}\textbf{Cost}\end{turn} & \begin{turn}{-90}\textbf{Energy}\end{turn} & \begin{turn}{-90}\textbf{Capacity\;}\end{turn} & \begin{turn}{-90}\textbf{Latency}\end{turn} & \multicolumn{1}{c|}{} &  & \multicolumn{1}{c|}{} & \multicolumn{1}{c|}{} \\ \hline
\cite{M1} & $D\mapsup C$ &  & $\checkmark$ & $\checkmark$ &  & ML supervised & \textit{1} & Experimental & Multimedia apps. \\ \hline
\cite{M2}  & \multirow{12}{*}{$D\mapsup E$} &  & $\checkmark$ &  & $\checkmark$ & (DRL)DQN & \textit{1} & Simulation & Real-time video   analytic \\ \cline{1-1} \cline{3-10} 
\cite{M3}  &  &  & $\checkmark$ &  & $\checkmark$ & (DRL)E2D & \textit{1} & Simulation & Video, smart home, and AI apps \\ \cline{1-1} \cline{3-10} 
\cite{M4}  &  &  & $\checkmark$ &  & $\checkmark$ & (MADRL)DDPG & \textit{n} & Simulation & UAV based application \\ \cline{1-1} \cline{3-10} 
\cite{M5}  &  &  & $\checkmark$ &  & $\checkmark$ & (MARL)Q-Learning & \textit{n} & Simulation & Generic user   applications \\ \cline{1-1} \cline{3-10} 
\cite{M6}  &  &  & $\checkmark$ &  & $\checkmark$ & (DRL)DQN & \textit{1} & Simulation & Generic user   application \\ \cline{1-1} \cline{3-10} 
\cite{M7}  &  &  & $\checkmark$ &  &  & (DRL)DQN & \textit{1} & Simulation & Generic user   applications \\ \cline{1-1} \cline{3-10} 
\cite{M8}  &  & $\checkmark$ & $\checkmark$ &  & $\checkmark$ & (DRL)DDPG+ TADPG & \textit{n} & Simulation & Generic user   applications \\ \cline{1-1} \cline{3-10} 
\cite{M9}  &  &  &  &  & $\checkmark$ & ML supervised & \textit{1} & Simulation & Resource-hungry IoT apps. \\ \cline{1-1} \cline{3-10} 
\cite{M10}  &  &  &  & $\checkmark$ & $\checkmark$ & (MADRL)DDPG & \textit{n} & Simulation & IIoT \\ \cline{1-1} \cline{3-10} 
\cite{M11}  &  &  &  & $\checkmark$ & $\checkmark$ & DL & \textit{1} & Simulation & Generic user applications \\ \cline{1-1} \cline{3-10} 
\cite{M12}  &  &  &  & $\checkmark$ & $\checkmark$ & (DRL)DQN \& DDPG & \textit{1} & Simulation & Generic user applications \\ \cline{1-1} \cline{3-10} 
\cite{M13}  &  &  &  &  & $\checkmark$ & (DRL)DDPG+ Optimization & \textit{1} & Simulation & Generic user applications \\ \hline

\cite{M14}  & \multirow{2}{*}{$D2D$, $D\mapsup E$} &  &  &  & $\checkmark$ & (DRL)DDPG & \textit{1} & Simulation & Resource-hungry   applications \\ \cline{1-1} \cline{3-10} 
\cite{M26}  &  &   & $\checkmark$&  & $\checkmark$ & (DRL)DQN & \textit{1} & Simulation & Generic user applications \\ \hline

\cite{M15}  &\multirow{2}{*}{$D\mapsup E$, $E\mapsup C$}  &  &  &  & $\checkmark$ & DL & \textit{1} & Simulation & IIoT\\ \cline{1-1} \cline{3-10}
\cite{ADM2}  &  &  &  &  & $\checkmark$ & (MARL)DDPG & \textit{n} & Simulation & IoT applications \\ \hline

\cite{M16}  & $C\mapsdown E$, $C\mapsdown D$ &  &  &  & $\checkmark$ & (MARL)Q-Learning & \textit{n} & Simulation & Generic user   applications \\  \hline
\cite{M17}  & \multirow{2}{*}{$D\mapsup F$} &  &  & $\checkmark$ & $\checkmark$ & ML supervised & \textit{1} & Experimental & IoT with ML jobs \\ \cline{1-1} \cline{3-10} 
\cite{M18}  &  &  & $\checkmark$ &  & $\checkmark$ & (DRL)DQN & \textit{1} & Simulation & Generic user   applications \\ \hline

\cite{ADM1}  & $D\mapsup F$, $F\mapsup C$ &  &$\checkmark$  &  & $\checkmark$ & (MADRL) DQN & \textit{n} & Simulation & IIoT applications \\ \hline

\cite{M19}  & $D\mapsup F$, $F\leftrightarrow F$, $F\mapsup C$ &  &  &  & $\checkmark$ & DL-unsupervised & \textit{1} & Simulation & Mobile applications \\ \hline

\cite{M20}  & \multirow{3}{*}{$V\mapsup E$} &  &  & $\checkmark$ &  & DL + Pareto   optimization & \textit{1} & Simulation & V2X applications \\ \cline{1-1} \cline{3-10}
\cite{M21}  &  &  & $\checkmark$ &  & $\checkmark$ & (DRL) DQN & \textit{1} & Simulation & V2X applications \\ \cline{1-1} \cline{3-10}
\cite{M22}  &  &  &  & $\checkmark$ &  & ML, MAB & \textit{1} & Emulation & V2X applications \\ \hline
\cite{M23}  & $V\leftrightarrow V$, $V\mapsup E $&  &  &  & $\checkmark$ & MAB & \textit{1} & Simulation & IoT application \\ \hline
\cite{M24}  & \multirow{2}{*}{$V\mapsup E$, $E\mapsup C$} & $\checkmark$ &  &  &  & (DRL) DQN & \textit{1} & Simulation & Generic user   applications \\ \cline{1-1} \cline{3-10} 
\cite{M25}  &  &  & $\checkmark$ &  & $\checkmark$ & (MARL)DDPG+LSTM & \textit{n} & Simulation & Payment application \\ \hline

\end{tabular}}

\end{table*}


A summary of ML-based offloading literature is shown in Table~\ref{tabel-5}. The comparisons are classified based on the offloading types in the federation. Like traditional optimization-based offloading, the commonly used metrics in the literature are cost, energy, capacity, and latency. The ML-based offloading methods are supervised ML, DL, RL, and DRL. Each ML algorithm has different characteristics, which are discussed in Table~\ref{tabel-4}.

\subsection{Traditional Optimization vs. Machine Learning} 
Three reasons why machine learning is required for offloading federated MEC systems are summarized in Table \ref{table-6}. First, a control plane module must make an immediate choice about offloading. Traditional optimization, with its high computational complexity and exhaustive searching, is not capable of meeting a control plane’s latency requirement. Second, monitoring dynamic MEC environments is not trivial and can introduce unknown information into the control plane module that is responsible for determining offloading policy. Third, modelling a heterogeneous MEC system precisely is challenging. Some researchers carried out  traditional optimization in federated offloading using a system snapshot. 

The papers \cite{M11, M13, ADM2, M26} employed ML to achieve fast offloading decisions in a complex federated system. These offloading decisions and resource allocations were modelled as mixed-integer nonlinear programming (MINLP) that would take a long time to solve by conventional optimization. Yang et al. \cite{M11} used DL approaches that solved the MINLP problem in near-real-time. DL also outperforms a conventional branch-and-bound algorithm in terms of system costs. A mobile device in an MEC system should take an online offloading decision in a complex and dynamic system which makes relaxation-based and local-search-based approaches to rerun in every change to the environment. These traditional optimization algorithms carry out exhaustive searching, which is not suitable for online decisions. Zhang et al. \cite{M26} extended a heuristic algorithm to the DQN, resulting in a fast-convergence algorithm suitable for real-time application offloading, and Huang et al. \cite{M13} proposed a Lyapunov-aided DRL framework to determine offloading policy in near-real-time with a near-optimum result compared to exhaustive searching approaches. 

Offloading in dynamic federated systems with unknown information was considered by proposing ML-based approaches in the papers \cite{M20, M25, M6, M10, M21}. Fan et al. \cite{M20} extended an SDN-controller with DL to learn a dynamic V2X system and carried out optimum offloading. This approach outperformed conventional traffic offloading (CTO), which uses heuristic algorithms, in terms of network throughput. Gao et al. \cite{M25} modelled offloading problem of V2X systems into Multi-Armed Bandit (MAB) and solved it by Probability-Based V2X Communication(PBVC) and adaptive learning-based task offloading (ALTO). Ale et al. \cite{M6} proposed DRL to address dynamic MEC systems for IoT. The current optimization techniques only take a snapshot of a system and cannot not address the dynamic environment. In their previous work Ale et al. \cite{MA1}, predicted traffic conditions and updated the cache by using DL. However, DL needs a large, labeled data-set to train models. 

Channel conditions, available communication, and computation resources change dynamically over time. Such changes may render some information unknown to IIoT agents which determine offloading policy. Guo et al. \cite{M10} used a multi-agent DDPG approach to tackle an offloading problem with some unknown or incomplete information. To ensure that a conventional algorithm, such as Greedy, works in this scenario, assumptions such as requiring agents to be aware of the channel and resource conditions in real-time, were made. In terms of the success rate in utilizing available channels, the results showed that MADDPG outperforms the Greedy algorithm. Zhaolong et al. \cite{M21} addressed offloading and resource allocation problems by using a DRL approach.  The proposed DRL approach had higher system utilities than a Greedy and a little lower than Brute-force. However, Brute-force carried out exhaustive searching, which is not suitable for a control plane. 

A heterogeneous federated system is difficult to model precisely, which make traditional offloading optimization difficult to implement. The papers \cite{M22, M14} used ML to carry out offloading in such a heterogeneous system. In Sonmez et al. \cite{M22}, the ML-based approach outperformed the Game-theory-based optimization in term of the success of tasks. Quality of experience (QoE)-aware task offloading in a Mobile Edge Network (MEN), which has heterogeneous computation and communication resources, is difficult to model for conventional optimization. He et al. \cite{M14} therefore proposed Double DDPG with which its learning agents could automatically update its model according to its experiences in interacting with the environment. This proposed method outperformed Greedy in term of latency.

\begin{table*}[!t]
\renewcommand{\arraystretch}{1.5}
\caption{Traditional Optimization vs. Machine Learning for Offloading in Federation.}
\label{table-6}
\begin{tabular}{|l|l|p{3.5cm}|p{3cm}|p{6.4cm}|}
\hline
\textbf{References} & \textbf{ML Approach} & \textbf{Traditional Optimization}                    & \textbf{Reason of using ML}                                  & \textbf{Conclusion}                                                                                    \\ \hline
\cite{M11}             & DL                   & Branch and bound                                      & \multirow{3}{*}{Computation complexity}                      & ML-based offloading has lower cost                                                                     \\ \cline{1-3} \cline{5-5} 
\cite{M26}             & DRL                  & Greedy and Heuristic  &                                                              & ML-based offloading has lowest   convergence time with better latency and energy usage \\ \cline{1-3} \cline{5-5}
\cite{M13}             & DRL                  & Relaxation-based, and   local-search-based approaches &                                                              & ML-based offloading has lowest convergence time with near-optimum result in term of computation rate \\ \hline \cline{5-5}

\cite{M20}              & DL                   & Heuristic CTO                                         & \multirow{5}{3cm}{Unknown information in   dynamic Environtment} & ML-based offloading has higher throughput                                                              \\ \cline{1-3} \cline{5-5} 
\cite{M25}          & MAB                  & Greedy                                                &                                                              & ML-based offloading has lower latency                                                                  \\ \cline{1-3} \cline{5-5} 
\cite{M6}            & DRL                  & Greedy                                                &                                                              & ML-based offloading has more completed task                                                            \\ \cline{1-3} \cline{5-5} 
\cite{M10}          & MADDPG               & Greedy                                                &                                                              & ML-based has lower latency and   has higher channel access success rate                                \\ \cline{1-3} \cline{5-5} 
\cite{M21}      & DQN                  & Greedy, Bruteforce                                    &                                                              & ML-based offloading has higher   system utilities than greedy and little bit lower than Bruteforce     \\ \hline
\cite{M22}     & ML, MAB              & Game theory optimization                              & \multirow{2}{*}{Heterogenous environment}                    & ML-based offloading has fewer failed task                                                              \\ \cline{1-3} \cline{5-5} 
\cite{M14}         & Double DDPG          & Greedy                                                &                                                              & ML-based offloading has lower latency                                                                  \\ \hline

\end{tabular}
\end{table*}

The references in Table \ref{table-6} do not specifically compare the traditional optimizations with the ML-based approaches. Most of them used model-free reinforcement learning approaches, such as DQN and DDPG, because these can directly adopt a model from the environment and do not need to provide the environment’s model to the learning agent. The Greedy algorithm is the preferred traditional algorithm because, with incomplete information from the environment, Greedy can still converge, although it may become stuck in local optima/minima. ML-based approaches can converge faster than traditional optimization with near-optimum results.

\section{Lessons Learned}

We categorize the lessons learned from this survey on the approaches that were used in the survey, such as traditional optimization and machine learning.

\subsection{Traditional Optimization-Based Offloading}
Some understanding comes from survey of papers on traditional optimization-based offloading which explore the basic idea of carrying out offloading in a cloud-edge-fog system.

\emph{1) Traffic offloading is a short-term solution to the dynamic arrival traffic rate, while capacity allocation is a long-term solution.} Traffic and task offloading in an MEC system can be categorized into control plane and management plane problems. On the arrival of traffic or a task, the control plane decides on offloading policy, leading to an objective such as minimizing latency. The control plane responds to arriving traffic in a matter of seconds. On the other hand, the management plane forecasts future traffic or task arrival rates based on historical data. The system’s capacity is then scaled to accommodate the predicted offloaded traffic. By integrating the control and management plane modules, it is possible to meet the arrival traffic or task’s latency requirements while allocating the fewest possible resources.

\emph{2) There are two offloading decisions to be made-- where to offload, and how much to offload.} An offloading decision could be a binary decision, which is a decision to offload or not, or a ratio-based offloading decision, which determines how much and where to offload tasks or traffic. Binary offloading is usually carried out by UEs, as UEs lack complete knowledge of external system resources. An UE measures its capacity to compute a task locally or to offload to external resources. Ratio-based offloading is carried out by network devices controlled by an orchestrator, which has global information to determine where and how much to offload. 

\emph{3) Hierarchical offloading--application offloading by UE and traffic/task offloading by network control plane.}
UEs carry out offloading to extend their computation capacity and extend their battery life since UEs are equipped with limited computation and battery capacity. UEs sense environment conditions such as signal strength, battery level and resources utilization (local information) to determine where a task is to be executed. Network devices offload their task to another network device with the least load to avoid overloading and minimize latency. These network devices could be a router, traffic dispatcher, MEC servers or fog servers with a data plane function. Network device offloading is determined by an orchestrator which has access to global system information.

\emph{4) Infrastructure capacity expands UE capacity, while non-infrastructure capacity extends UE connections.} Traffic or tasks can be offloaded to infrastructure and/or non-infrastructure. Infrastructure comprises all entities that belong to providers or organizations, such as base stations, MEC servers, fog nodes, and the cloud. Such infrastructure entities are used to extend a mobile device and UEs’ capacity. Some areas may, however, not be covered by infrastructure entities. UEs offload traffic or tasks to another UE or mobile device (non-infrastructure) in such an area, called an opportunistic network. A UE or mobile device can share its computing capacity as a server or share its communication capability as a relay to infrastructure in an opportunistic network.

\emph{5) Horizontal offloading boosts east-west traffic while keeping traffic at lower tier with low-latency services. Vertical offloading, on the other hand, minimizes capacity allocation and simplifies management by centralizing the upper tier.} A federation of cloud-edge-fog is a hierarchical system in which cloud is at the top, edge is in the middle, and fog is at the bottom. Furthermore, each tier may include some providers. This system has two offloading directions, which can be bottom-up (vertical) or east-west (horizontal).

Vertical offloading occurs between the customer and provider or between tiers within a provider, such as offloading in two-tier MEC architecture. In customer-provider (upward) vertical offloading, a customer will be charged for each resource used, and minimizing costs will be of concern in such cases. On the other hand, an upper-tier provider may offload a service to a lower-tier provider in order to meet required latency (downward offloading). The upper tier provider pays an incentive to a lower tier provider for every served task. The upper-tier provider minimizes offloading costs while maintaining the required latency.

Vertical offloading can also be carry out within a provider and is typically used to move tasks from a lower, more dispersed, tier to a higher, more centralized, tier with larger capacity. The upper tier provides greater coverage and shares its capacity with a couple bottom-tier sites in order to handle high arrival traffic or task rates that would overwhelm some bottom-tier sites.

Horizontal offloading is carried out to distribute traffic or tasks to the same tier first, rather than offloading them to a higher tier. Keeping tasks on the bottom tier, such as fog or edge, might help reduce communication latency due to their proximity to the UE or MD. While horizontal offloading keeps traffic on the bottom tier, the decision to offload must be made on the neighboring side, otherwise the traffic will encounter prolonged communication latency, and a  trade-off between minimizing computing and communication latency takes place.

\emph{6) Service chaining offloading is unavoidable, given that modern services are composed of multiple microservices.} In task offloading, a task can be a single task or consists several sub-tasks. Full offloading is carried out for a single task which can be processed at local device (UE) or remotely (cloud, edge or fog). While in partial offloading, some sub-tasks can be processed locally, and others are offloaded to a remote server.  Sub-task synchronization may be needed if some sub-tasks depend on other sub-task output.

\emph{7) Different providers have different objectives, such as minimizing cost or latency, and they might compete with each other.} Optimizing offloading decisions in a federated system with some providers is challenging because each provider has its own objective. A collaborative offloading approach by adopting game theory and analysing Nash equilibrium is carried out. In the Nash equilibrium, the offloading strategy of each provider is optimal when taking another providers’ decisions into account. Each provider ends up winning since everyone gets the result they expect.

\subsection{Machine Learning-Based Offloading}
Some insights were gained from a survey of ML-based offloading in a cloud-edge-fog federation. An offloading decision in a cloud-edge-fog federation is made by the control plane and applied in the data plane of the networking devices. This control plane decision must be carried out quickly (fast response time). Making a quick decision in a federated system with high complexity is very challenging. Traditional optimization, which uses exhaustive searching, may violate the latency requirements of a control plane decision. ML-based offloading is a promising method that automatically maps a given system settings to arrive at the best offloading decision.

\emph{1) Offline and online learning of ML-based offloading optimization approaches.}
Unsupervised-based ML is used to predict future system conditions by using earlier/older data. The label of the data is not essential in unsupervised learning. The predicted results are used by the control plane to make the best offloading decisions. Unlike unsupervised learning, supervised learning trains the model directly, based on collected data that an expert has labelled. The label is an offloading decision that leads to an objective such as minimizing delay or costs for a given current system condition.

While supervised learning can derive an optimal solution, deriving a well-labelled data-set is not easy; extensive monitoring is required. Some monitoring mechanisms that carry out costly broadcast data are required to obtain information in distributed edge-fog systems. A label of data must be updated frequently in a dynamic system because an offloading decision must be carried out quickly.

Both unsupervised and supervised learning are categorized as offline learning because they learn from previous data, and not directly from the environment. Reinforcement learning is one of the ML-based approaches which learns directly from the environment to determine the best action. In a cloud-edge-fog federation, an offloading decision is produced by a learning agent. In the beginning, the agent will perform poorly; it will remember previous successful actions taken in a given environment state condition and forget failure action. With such trial-and-error attempts, the agent will improve after several attempts.

\emph{2) Machine learning approaches are a panacea for optimizing offloading decisions with some missing information.} Some providers in a federation may hide some information from others.  However, traditional optimization requires all information to calculate optimal offloading. Some researchers made assumptions about such hidden information because the calculation could otherwise not have been carried out. By contrast, ML-based offloading will map any given input that may also be incomplete to determine the  best offloading decision. This learning process can be carried out with incomplete information. 

\emph{3) Retraining is more appropriate than recalculation for control plane problems, such as offloading optimization.} In terms of the cost to obtain an optimal decision, traditional optimization recalculates the decision for each new given input before deriving a new offloading decision. The ML-based solution will retrain the model to obtain an optimal decision. Without retraining process, the model could still come up with  a sub-optimal result.  ML-based approaches can thus decide quickly, without waiting for the training process to be completed; and ML models can also be reused and transferred.

\emph{4) Federated systems are a type of multi-agent environment.} A learning agent, a control plane module, can be a single agent placed in a central location or multi-agent distributed over some areas. A single agent determines the offloading decisions for all devices in a federation. An agent’s model is trained by a centralized data-set. The size of the federation system will affect the data-set’s dimensions and raise a scalability problem because of the very large dimensions of the observations and decisions. Single-agent learning is unrealistic because a federation consists of many providers who have different offloading policies. Multi-agent learning is suitable in a federated system for two reasons. The first reason is scalability. A provider may have agents in some areas which produce offloading decisions based on local observations. Having multi-agents which calculate offloading decisions in parallel can reduce convergence times. The second reason is the federation itself because the federation is a kind of multi-agent environment where each agent belongs to each provider.

\emph{5) Federated learning sharing  model weights is a promising approach to obtain global optimum offloading decision with low communication cost.} A cloud-edge-fog federation may consist of millions of devices scattered over large area.  A learning agent that produces offloading decisions can be trained by a centralized data-set with global information, or trained by a local dataset consisting of local information; being trained by local dataset results in a local optimum offloading decision. While using a centralized data-set with global information results in a globally optimum offloading decision, but generates extensive communication costs.

\subsection{Traditional vs. Machine Learning Based Offloading}

\emph{1) While traditional optimization determines the optimum offloading action based on a snapshot of the system, machine learning-based offloading would make use of continuous system information.} Because a system snapshot may not accurately represent future system behavior, offloading becomes obsolete in traditional optimization. ML-based techniques, particularly RL-based offloading, can be trained on batches of collected data without waiting for complete or a large amount of collected data from environment. \\

\emph{2) Traditional optimization techniques are suitable for management plane problems, while machine learning-based approaches are best suited for control plane problems.} A management plane, which is in control of resource allocation in a federated system, makes decisions in minutes or hours and so generates long-term solutions for hot-spot traffic. A management plane problem is well-suited to traditional optimization, which has a long decision time. While the short-term solution to hot-spot traffic is offloading, which is part of the control plane problem. Offloading using machine learning has a short decision time since it produces a sub-optimal solution after only a few training processes, making it appropriate for the control plane problem.\\

\emph{3) Transitioning from traditional optimization to machine learning-based approaches can minimise assumptions of unknown information in modeling federated systems.} Channel, network, and server settings are difficult for a learning agent to obtain completely. Certain pieces of information, such as the relationship between tasks and the processing capacity required to do those tasks, may be unknown to the learning agent. To deal with unknown information in traditional optimization, some researchers make assumptions such as that the computing and networking capacity is homogeneous and able to collect system information completely. Because machine learning-based techniques can map any input to a desired output, they can be utilized to minimize the assumptions of unknown information.

\section{Research Opportunities and Challenges}

\subsection{Research Opportunities}

\emph{1) Fog-Fog Federation.}
With development of fog computing, creates several benefits for the application developers, applications, and different industries by distributing functions \cite{C1}.
A fog-fog federation helps devices monitor, process, analyse, react, and distribute computation, communication, storage, control, and decision-making closer to users. However, such a federation also results in challenges for individual fogs. When fogs are closer to each other than to edge or cloud, the federation between fogs allows them to enhance their data aggregation, processing, and storage capabilities, and requires cooperation between these fogs to ensure the proper coordination for the necessary interactions. 

\emph{2) V2X.}
In the past few years, Internet use has continued to increase with the development of advanced technologies. The gradual increase of smart vehicle applications has produced computation-intensive tasks for vehicles, and thus, the internet of vehicles (IoV) improves  traffic conditions \cite{C5}.  
However, these vehicles are independently unable to meet the demands of their limited computing resources. Vehicle-to-everything (V2X) communication is an emerging technology that supports vehicles to offload their tasks across vehicles \cite{C6}.
With vehicle-to-infrastructure communication (V2I), a vehicle can offload to infrastructures such as RSU, edge, or cloud, and with vehicle-to-vehicle communication (V2V), one vehicle can offload its tasks to other vehicles. As an alternative, this technology also facilitates multiple vehicles forming a fog by sharing their resources, popularly known as a vehicular-fog, to provide services to others.

\emph{3) Mobility of a Vehicular-Fog.}
Intelligent transport systems (ITS) \cite{C7} exchange information for safe V2V and V2I communication. In a V2V environment, vehicles communicate directly with each other and with services that support safe driving and provide information. However, in a dynamic environment, some applications of the moving nodes require high computing power, and the computational resources of each vehicle may not be able meet such a requirement. We address this issue of V2I communication by using MECs or RSUs that are closer to the vehicles. However, in a certain dynamic environment, vehicles can move out of communication range during task offloading. In such a case, either the task cannot be offloaded to the infrastructure, or, if offloaded, the vehicle cannot receive the results. In an environment where vehicles are on the move, the rate of movement of  vehicles is usually fast, and the change of topology  intense,  and it is here where  V2V task offloading  is a matter that needs to be  investigated. Recent vehicular-fog research has focused mostly on the static vehicle scenario \cite{B25},
whereas in a vehicular-fog set-up vehicles are mostly managed by a fog manager like RSUs, which is part of the infrastructure. In a dynamic vehicular-fog with mobility management of the federation is a matter of concern.

\emph{4) Scaling.}
Auto-scaling \cite{C9} can be classified into different categories. First, manual scaling, where we specify only the maximum, minimum, or desired capacity changes to auto-scaling groups, and auto-scaling maintains the instances with updated capacity. Second, scaling based on a schedule, where one can scale an application ahead of known load changes. For example, on some particular day, in peak loads or on a limited offer, one can scale an application based on scheduled scaling in such cases. Third, for scaling based on demand or dynamic or reactive scaling, resources are adjusted in real-time based on the number of incoming requests. Finally, predictive scaling predicts future arrival traffic rates by learning past arrival traffic information, and the learning outcomes are then used to make scaling decisions. In a federated system, service providers can scale resources up or down by adopting different scaling methods. It is also essential for a service provider to make decisions based on different performance metrics, whether to scale up its resources to accommodate more incoming requests or offload the request to others. So, it remains challenging to decide when the resources need to be scaled up to avoid offloading and when to be encouraged to offload to avoid scaling \cite{E3}.

\emph{5) Centralized vs. Distributed Federation.}
Various factors affect a federation, such as the services available from service providers, the type of services, their capacity and capabilities, their geographical location, number of customers, type of customers, etc. 
A federation between the service provider can be centralized or distributed. 
A centralized federation has a single federation manager between multiple federated entities, and that manager manages the federation. In such cases, there is a joint federation agreement between all the entities, based on which offloading decisions will be taken \cite{C4}. 
In a distributed scenario, a federation is formed between two individual entities or a group of entities of a system. In such cases, a separate agreement is made based on what communications take place.

\emph{6) Resource Allocation.}
When the number of resources in a system is large, some may remain underutilized.
If the number of resources is small, offloading may be triggered too often. 
Hence, one of the key challenges of offloading is to determine the right amount of resources required at the location where the tasks will be executed, otherwise, after offloading, if there is any shortage of resources, tasks will be offloaded further away \cite{E4}. 
This may increase the communication latency as well as that an increase in the number of offloading hops may trigger a breach of data privacy \cite{E5}. 
Again, most applications and services in the system that require intensive computation and high processing are incompatible with devices because of their limited resources. 

\emph{7) Energy Consumption.}
Although task offloading is largely inevitable in a federated system, it is still a high energy consuming process. One of the challenges is to estimate the energy consumed in communication activities of task offloading so as to make task offloading efficient \cite{E6}.
As a result, it is sometimes a challenge whether to offload or not. An efficient energy estimation model would help to decide whether or not to perform task offloading, based on the energy cost of the communication activities.

\emph{8) Task Offloading in Different Application Scenarios.}
Task offloading can take place at different locations of different federated systems, depending on the type of service required, and based on different criteria. This section is an overview of different application scenarios where task offloading has recently played a key role. In ITS, automatic traffic monitoring and management systems \cite{E7}, edges and RSUs can assist drivers by providing traffic updates, emergency alerts, etc. In fog-to-fog offloading scenarios, one vehicle can assist another by caching data if required  \cite{C5}. 
Emergency help alert mobile cloud (E-HAMC) can provide a quick way of notifying the relevant emergency authorities by utilizing the services of fog for offloading and pre-processing purposes \cite{E9}. When an alert message is sent, these services can automatically transmit the location of an incident and the emergency contact information. New offloading schemes can improve privacy levels, reduce computation latency and save the energy consumption of healthcare IoT devices \cite{E10}. 
Edge Computing also offers intriguing possibilities for smart agriculture \cite{E11}. For example, sustainable water management is a common issue at farm level. By offloading sensible data from the sensors devices to the edge server, appropriate action can be taken.

\subsection{Research Challenges}

\emph{1) Interoperability.}
Interoperability is closely related to both standards and lock-in \cite{C2}. 
Internet service providers use multiple networks so that the failure of a single provider will not disrupt communications entirely. Here we will focus on interoperability between cloud, edge, and fog providers. In a federated system, an application’s execution can be carried out with its components spread over different service providers. From an architectural perspective, appropriate signalling, data, and control interfaces are needed to ensure interoperability at the architectural level, or more precisely, to support an application’s life-cycle, the control interfaces are needed for interactions between the different domains.

\emph{2) Service Level Agreement.}
Every user wants assurances that their service provider will remain reliable because service interruptions can cause significantly financial harm. Service Level Agreements (SLAs) \cite{C8} are contractual agreements for certain levels of reliability, which would then be compensated in various ways if there was any breach of the contract. The same kind of agreement is applicable between the service providers who federate with each other to provide services to their respective subscribers. There must be a contract, a federation agreement, to provide a certain level of reliability. Such provisions may include monetary compensation if the level of service offered is below the contractually specified level.

\emph{3) Redundancy.}
Redundancy is crucial in numerous scenarios to ensure the system's high availability. Obviously redundancy is more critical at the cloud level as compared to edge and fog, as a cloud has multiple data centers, redundant networking, backup power, data backup plans and other redundancy resources.

\emph{4) Fault Tolerance.}
In a federated system, offloading gradually becomes automated, and where heterogeneous entities are involved, the risks of failure increases. Some common examples are connectivity failure, use of faulty devices, communication delays, etc.  Offloading processes must then be robust and must be capable of not only detecting but also of handling faults on time. The accuracy and timeliness of the fault detection algorithm to detect the faults are thus of significant importance.

\emph{5) Security.}
In a federated architecture, multiple systems communicate with each other, and when offloading occurs, there is a risk of data theft and misuse. The misuse of data can be a serious threat to security systems such as of the military, healthcare, etc. It may also compromise the privacy of individuals. Hence, efficient and robust data security measures would be required so that offloading decisions are precise because security breaches are something that may not be publicly disclosed by service providers, unless compelled to do so by particular regulations \cite{C11}.

Cloud, edge and fog can be federated in several different ways in which a subscriber can move from using the services of one (cloud, edge or fog) to using the services of another. There is then a need for authentication when a subscriber moves between two entities that are federated, directly or indirectly via some hops in-between. This leads to 3rd-party  \cite{C12} and 4th-party authentication.

\emph{6) Geo-Diversity.}
The location and geographical diversity of a service provider might be of concern to some users. Compared to centralized cloud systems, widely spread geographical distribution of fogs and edges can be considered one of the key enablers of the Internet of Things (IoT) and big data applications \cite{C10}. These offer low latency and location awareness due to the proximity of the computing devices.

\emph{7) Reliability.}
The reliability track record of a service provider are just as crucial as contractual guarantees. Big cloud providers are likely to have significantly better reliability than relatively small, self-maintained IT infrastructures, as they have massive computing capabilities. Edge and fog computing systems are closer to a user and improve user experience by providing low latency and highly efficient computing. When computationally intensive components are offloaded to edge servers or distributed to fog nodes, various constraints such as power limitations, limited computing resources, inevitable server failure, etc. come into play. In such a scenario, how is the reliability of  offloaded computing \cite{C3} to be guaranteed? How then does one find an appropriate offloading point that can guarantee completing a task at low cost, with minimal energy consumption for communication? What is achievable minimal latency for the completion of the task?

\emph{8) Performance.}
In federated systems, many customers may share common physical computer hardware and network infrastructure. However, sharing can also cause performance problems. As providers use statistical multiplexing, excessive levels of over-subscription may degrade services. Poor resource scheduling and poor management could also degrade performance, even if there is no over-subscription. If any service provider in a federated system misrepresents their available capacity or capability, this two may cause performance degradation. 


\section{Conclusion}

Network communication relies on the coexistence of a variety of architectures of different service. The coexistence of such distinct architectures and complementary technologies opens up new issues in resource, latency and storage limitation, type of services, etc., which can emerge with federation architectures and cannot be addressed individually.

We have discussed various federation architectures for cloud, edge, and fog systems.  These each have their own tier within a federation, as top, middle, and bottom tiers. Such a federation can be classified as vertical, horizontal, or hybrid. Horizontal federation is that between providers of the same tier, vertical federation is a federation at different levels that can result in a 2-tier or 3-tier architecture, and a hybrid federation is the term used to describe the combination of vertical and horizontal federations.

We also give an overview of the various offloading techniques in such a cloud-edge-fog federated system and classify them according to the federation relationship and direction of offloading, i.e. horizontal offloading occurs in a horizontal federation, and so on.  Most of the works we have considered have focused exclusively on vertical offloading within a federation. Horizontal offloading should also be considered, as there are multiple providers within the same federation tier that can provide resources.

We also reviewed literature on various recently proposed offloading approaches, categorizing them as traditional optimization and machine learning-based approaches. In a federated system, the high dimension and dynamic input with unknown input parameters complicate the calculation of offloading decisions. Offloading decisions in the control plane module must be made quickly. The traditional optimization approach, which relies on exhaustive searching, may violate a control plane’s latency requirements. Machine learning-based approaches that map any input parameters (even with unknown ones) for a desired output have emerged as a solution to the limitations of traditional optimization. The machine learning model can also be used to generate immediate offloading decisions without waiting for all training processes to be finalized, which would result in an optimal solution. Because reinforcement learning approaches can derive an offloading decision directly from the environment without requiring a well-labelled data set, they become the most preferred approach.

Finally, we discussed some future research directions for such offloading scenarios and highlighted some key challenges associated with the task offloading.


%



\section*{Acknowledgment}
This work was supported by the Ministry of Science and Technology (MOST), Taiwan under Grant 109-2221-E-011-104-MY3.

\ifCLASSOPTIONcaptionsoff
  \newpage
\fi

\end{document}